\documentclass{vldb}

\usepackage{enumerate}
\usepackage[shortlabels]{enumitem}

\usepackage{hyperref}

\usepackage{algorithmicx}
\usepackage{algorithm}
\usepackage{algpseudocode}
\usepackage{balance}
\usepackage{booktabs}
\usepackage{graphicx}
\usepackage{nth}
\usepackage{pervasives}
\usepackage{protocoldiagrams}
\usepackage{protocol}
\usepackage{subcaption}
\usepackage{tikz}
\usetikzlibrary{backgrounds}
\usetikzlibrary{calc}
\usetikzlibrary{decorations.pathreplacing}
\usetikzlibrary{positioning}

\ifdefined\turnontechreport
  \toggletrue{techreportenabled}
\else
  \togglefalse{techreportenabled}
\fi
\toggletrue{techreportenabled}

\newcommand{\thetitle}{Matchmaker Paxos: A Reconfigurable Consensus Protocol\iftoggle{techreportenabled}{ [Technical Report]}{}}

\algtext*{EndIf}
\algtext*{EndFor}

\algblockdefx[Upon]{Upon}{EndUpon}
  [1]{\textbf{upon} #1 \textbf{do}}
  {end}
\algtext*{EndUpon}

\algnewcommand{\algorithmicstate}{\textbf{State:}}
\algnewcommand{\GlobalState}{\item[\algorithmicstate]}

\vldbTitle{\thetitle}
\vldbAuthors{%
  Michael Whittaker,
  Neil Giridharan,
  Adriana Szekeres,
  Joseph M. Hellerstein,
  Heidi Howard,
  Faisal Nawab, and
  Ion Stoica%
}
\vldbDOI{https://doi.org/10.14778/xxxxxxx.xxxxxxx}
\vldbVolume{13}
\vldbNumber{xxx}
\vldbYear{2020}

\begin{document}
\title{\thetitle}
\numberofauthors{7}

\author{
  \centering
  Michael Whittaker$^1$,
  Neil Giridharan$^1$,
  Adriana Szekeres$^2$,
  Joseph M. Hellerstein$^1$,
  Heidi Howard$^3$,
  Faisal Nawab$^4$,
  Ion Stoica$^1$
  \\[12pt]
  $^1$ University of California, Berkeley,
  $^2$ University of Washington,
  $^3$ University of Cambridge,
  $^4$ University of California, Santa Cruz
  \\[12pt]
}


\maketitle

{\begin{abstract}
  State machine replication protocols, like MultiPaxos and Raft, are at the
  heart of nearly every strongly consistent distributed database. To tolerate
  machine failures, these protocols must replace failed machines with live
  machines, a process known as reconfiguration. Reconfiguration has become
  increasingly important over time as the need for frequent reconfiguration has
  grown. Despite this, reconfiguration has largely been neglected in the
  literature.
  %
  In this paper, we present Matchmaker Paxos and Matchmaker MultiPaxos, a
  reconfigurable consensus and state machine replication protocol respectively.
  Our protocols can perform a reconfiguration with little to no impact on the
  latency or throughput of command processing; they can perform a
  reconfiguration in one round trip (theoretically) and a few milliseconds
  (empirically); they provide a number of theoretical insights; and they
  present a framework that can be generalized to other replication protocols in
  a way that previous reconfiguration techniques can not. We provide proofs of
  correctness for the protocols and optimizations, and present empirical
  results from an open source implementation.

  \begin{techreport}

    This version of the paper is a technical report that includes some additional
    information that is not present in our main publication. Additional text
    (like this) is annotated with a red bar along its left side.
  \end{techreport}
\end{abstract}
}
{\section{Introduction}
Strongly consistent distributed databases---like
Spanner~\cite{corbett2013spanner}, CockroachDB~\cite{cockroach2020}, and
FoundationDB~\cite{chrysafis2019foundationdb}---are becoming increasingly
adopted in industry and increasingly studied in academia. All of these
distributed databases rely on some state machine replication protocol, like
MultiPaxos~\cite{lamport1998part} or Raft~\cite{ongaro2014consensus}, to keep
multiple replicas of their data in sync. To ensure system availability, these
replication protocols must be able to dynamically replace failed machines with
live machines, a process known as reconfiguration.

Reconfiguration is an old problem, but recent computing trends make it a
particularly timely topic of research.
Historically, state machine replication protocols were deployed on a fixed set
of machines, and reconfiguration was used only to replace failed nodes with live
nodes -- an infrequent occurrence. Recently however, systems have become
increasingly elastic, and the need for frequent reconfiguration has grown.
These elastic systems don't just perform reconfigurations \emph{reactively}
when machines fail; they reconfigure \emph{proactively}.
For example, cloud databases can proactively request more resources to handle
workload spikes, and orchestration tools like Kubernetes~\cite{kubernetes2020}
are making it easier to build these types of elastic systems.
Similarly, in environments with short-lived cloud instances---as with
serverless computing and spot instances---and in mobile edge and Internet of
Things settings, protocols must adapt to a changing set of machines much more
frequently.

Despite the (increasing) importance of reconfiguration, it has largely been
neglected by current academic literature. Researchers have invented dozens of
state machine replication protocols, yet many papers either discuss
reconfiguration briefly with no evaluation~\cite{moraru2013there,
ports2015designing, rystsov2018caspaxos, rizvi2017canopus}, propose
theoretically safe but inefficient reconfiguration
protocols~\cite{lamport2005generalized, liskov2012viewstamped}, or do not
discuss reconfiguration at all~\cite{lamport2006fast, mao2008mencius,
marandi2010ring, biely2012s, arun2017speeding}.
%
%
In recent years, state-of-the-art databases---like MDCC~\cite{kraska2013mdcc},
Janus~\cite{mu2016consolidating}, and TAPIR~\cite{zhang2018building}---have
begun to adopt more sophisticated replication protocols, but these are exactly
the protocols for which there are not well-established, high performance techniques
for reconfiguration.
Inventing new reconfiguration protocols for these new databases is challenging.
Reconfiguration protocols are notoriously complicated and hard to get right. In
one recent anecdote, Raft's authors proposed a ``simpler single-server''
reconfiguration protocol to replace their original, more complicated protocol,
but this simpler protocol turned out to be unsafe~\cite{ongaro2020bug}.

In this paper, we present Matchmaker Paxos and Matchmaker MultiPaxos, a
reconfigurable consensus protocol and state machine replication protocol
respectively. Our protocols have the following desirable properties:

    \textbf{Little to No Performance Degradation.}
    Matchmaker MultiPaxos can perform a reconfiguration without significantly
    degrading the throughput or latency of processing client commands. For
    example, we show that reconfiguration has less than a 2\% effect on median
    and standard deviation latency measurements (\secref{Evaluation}).

    \textbf{Quick Reconfiguration.}
    Matchmaker MultiPaxos can perform a reconfiguration quickly. Theoretically,
    reconfiguring to a new set of machines takes one round trip of
    communication (\secref{MatchmakerMultiPaxos}). Empirically, this requires
    only a few milliseconds within a single data center (\secref{Evaluation}).

    \textbf{Theoretical Insights.}
    Matchmaker Paxos offers theoretical insights into many existing protocols
    (\secref{Generality}). It extends Vertical
    Paxos~\cite{lamport2009vertical}, it is the first protocol to achieve the
    theoretical lower bound on Fast Paxos~\cite{lamport2006fast} quorum sizes,
    and it corrects errors in DPaxos~\cite{nawab2018dpaxos}.

    \textbf{Generality.}
    Reconfiguration protocols are hard to invent, so ideally we would re-use
    existing protocols whenever possible. MultiPaxos' horizontal
    reconfiguration~\cite{lamport2010reconfiguring} is arguably the
    most popular reconfiguration technique, but it makes assumptions that make
    it incompatible with many existing replication protocols.  Matchmaker Paxos
    does not makes these restrictive assumptions, and can be more easily used
    by other replication protocols (\secref{Generality}).

    \textbf{Proven Safe.}
    We describe Matchmaker Paxos and Matchmaker MultiPaxos precisely and prove
    that both are safe (Sections \ref{sec:MatchmakerPaxos},
    \ref{sec:MatchmakerMultiPaxos}, \ref{sec:RetiringOldConfigurations},
    \ref{sec:ReconfiguringMatchmakers}). Unfortunately, this is not often done
    for reconfiguration protocols~\cite{moraru2013proof, ports2015designing,
    rystsov2018caspaxos, rizvi2017canopus}.

In a nutshell, our protocols work by leveraging two key design ideas. The first is to \emph{decouple reconfiguration} from the standard processing path.
%
Many replication protocols~\cite{lamport2010reconfiguring, ongaro2014consensus}
have nodes that are responsible for both processing commands and for
orchestrating reconfigurations.
By contrast, Matchmaker Paxos introduces a set of distinguished matchmaker nodes that are solely
responsible for managing reconfigurations. This decoupling, along with a number
of novel protocol optimizations, allow us to perform reconfiguration quickly in the background, without degrading performance.

The second design point is to \emph{reconfigure across rounds, not commands}.
Replication protocols based on classical MultiPaxos assume a totally ordered log of chosen commands, and reconfigure across log entries: each log entry is handled by a single configuration.
Matchmaker Paxos instead works \emph{across rounds of consensus},
as proposed by Vertical Paxos~\cite{lamport2009vertical}. Different Paxos rounds---even for the same command---can use
different configurations. Many replication protocols in fact have no sequential log, but almost every
one has some form of rounds (a.k.a.\ terms, epochs, views).
Matchmaker Paxos gets its generality from this design point, and goes beyond Vertical Paxos in terms of simplicity and efficiency (\secref{Generality}).

}
{\section{Background}\seclabel{Background}

\subsection{System Model} Throughout the paper, we assume an asynchronous
network model in which messages can be arbitrarily dropped, delayed, and
reordered. We assume machines can fail by crashing but do not act maliciously;
i.e., we do not consider Byzantine failures. We assume that machines operate
at arbitrary speeds, and we do not assume clock synchronization.  We assume a
discovery service that nodes can use to find each other, but do not require
that this service be strongly consistent. A node can safely communicate
with outdated nodes. A system like DNS would suffice. Every protocol discussed in
this paper assumes that at most $f$ machines will fail for some configurable
$f$.

\subsection{Paxos}
A \defword{consensus protocol} is a protocol that selects a single value from a
set of proposed values. \defword{Paxos}~\cite{lamport1998part,
lamport2001paxos} is one of the oldest and most popular consensus protocols.
We assume that the reader is familiar with Paxos, but we review the parts of
the protocol that are most important to understand for the rest of the paper.

A Paxos deployment that tolerates $f$ faults consists of an arbitrary number of
clients, $f+1$ nodes called \defword{proposers}, and $2f+1$ nodes called
\defword{acceptors}, as illustrated in \figref{PaxosBackgroundDiagram}.
To reach consensus on a value, an execution of Paxos is divided into a number
of rounds, each round having two phases: Phase 1 and Phase 2. Every round is
orchestrated by a single pre-determined proposer.
The set of rounds can be any unbounded totally ordered set
with a least element. It is common to let the set of rounds be the set of
lexicographically ordered integer pairs $(r, id)$ where $r$ is an integer
and $id$ is a proposer id, where a proposer is responsible for executing every
round that contains its id.
%

{
\tikzstyle{proc}=[draw, circle, thick, inner sep=2pt]
\tikzstyle{client}=[proc, fill=clientcolor!25]
\tikzstyle{proposer}=[proc, fill=proposercolor!25]
\tikzstyle{acceptor}=[proc, fill=acceptorcolor!25]

\tikzstyle{proclabel}=[inner sep=0pt, align=center, font=\small]

\tikzstyle{comm}=[-latex, thick]
\tikzstyle{commnum}=[fill=white, inner sep=0pt]

\begin{figure}[ht]
  \centering
  \begin{subfigure}[b]{0.45\columnwidth}
    \centering
    \begin{tikzpicture}[xscale=1.5]
      \node[client] (c1) at (0, 2) {$c_1$};
      \node[client] (c2) at (0, 1) {$c_2$};
      \node[client] (c3) at (0, 0) {$c_3$};
      \node[proposer] (p1) at (1, 1.5) {$p_1$};
      \node[proposer] (p2) at (1, 0.5) {$p_2$};
      \node[acceptor] (a1) at (2, 2) {$a_1$};
      \node[acceptor] (a2) at (2, 1) {$a_2$};
      \node[acceptor] (a3) at (2, 0) {$a_3$};

      \node[proclabel] (clients) at (0, 3) {Clients};
      \node[proclabel] (proposers) at (1, 3) {$f+1$\\Proposers};
      \node[proclabel] (acceptors) at (2, 3) {$2f+1$\\Acceptors};
      \halffill{clients}{clientcolor!25}
      \quarterfill{proposers}{proposercolor!25}
      \quarterfill{acceptors}{acceptorcolor!25}

      \draw[comm] (c1) to node[commnum]{1} (p1);
      \draw[comm, bend left=15] (p1) to node[commnum]{2} (a1);
      \draw[comm, bend left=15] (p1) to node[commnum]{2} (a2);
      \draw[comm, bend left=15] (a1) to node[commnum]{3} (p1);
      \draw[comm, bend left=15] (a2) to node[commnum]{3} (p1);
    \end{tikzpicture}%
    \caption{Phase 1}\figlabel{PaxosBackgroundDiagramPhase1}
  \end{subfigure}\hspace{0.05\columnwidth}
  \begin{subfigure}[b]{0.45\columnwidth}
    \centering
    \begin{tikzpicture}[xscale=1.5]
      \node[client] (c1) at (0, 2) {$c_1$};
      \node[client] (c2) at (0, 1) {$c_2$};
      \node[client] (c3) at (0, 0) {$c_3$};
      \node[proposer] (p1) at (1, 1.5) {$p_1$};
      \node[proposer] (p2) at (1, 0.5) {$p_2$};
      \node[acceptor] (a1) at (2, 2) {$a_1$};
      \node[acceptor] (a2) at (2, 1) {$a_2$};
      \node[acceptor] (a3) at (2, 0) {$a_3$};

      \node[proclabel] (clients) at (0, 3) {Clients};
      \node[proclabel] (proposers) at (1, 3) {$f+1$\\Proposers};
      \node[proclabel] (acceptors) at (2, 3) {$2f+1$\\Acceptors};
      \halffill{clients}{clientcolor!25}
      \quarterfill{proposers}{proposercolor!25}
      \quarterfill{acceptors}{acceptorcolor!25}

      \draw[comm, bend left=15] (p1) to node[commnum]{4} (a1);
      \draw[comm, bend left=15] (p1) to node[commnum]{4} (a2);
      \draw[comm, bend left=15] (a1) to node[commnum]{5} (p1);
      \draw[comm, bend left=15] (a2) to node[commnum]{5} (p1);
      \draw[comm] (p1) to node[commnum]{6} (c1);
    \end{tikzpicture}
    \caption{Phase 2}\figlabel{PaxosBackgroundDiagramPhase2}
  \end{subfigure}
  \caption{Paxos communication diagram ($f=1$).}%
  \figlabel{PaxosBackgroundDiagram}
\end{figure}
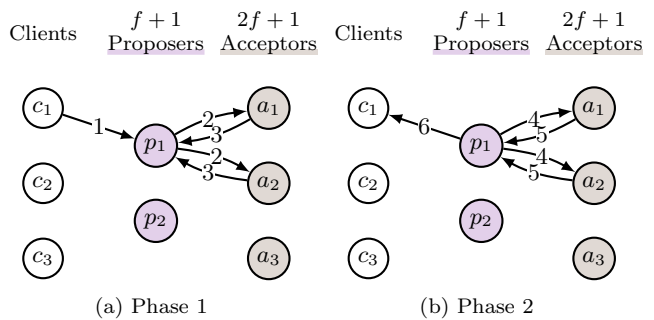}

When a proposer executes a round, say round $i$, it attempts to get some value
$x$ chosen in that round. Paxos is a consensus protocol, so it must only choose
a single value. Thus, Paxos must ensure that if a value $x$ is chosen in round
$i$, then no other value besides $x$ can ever be chosen in any round less than
$i$. This is the purpose of Paxos' two phases. In Phase 1 of round $i$, the
proposer contacts the acceptors to (a) learn of any value that may have already
been chosen in any round less than $i$ and (b) prevent any new values from
being chosen in any round less than $i$. In Phase 2, the proposer proposes a
value to the acceptors, and the acceptors vote on whether or not to choose it.
In Phase 2, the proposer is careful to only propose a value $x$ if it learned
through Phase 1 that no other value has been or will be chosen in a previous
round.

More concretely, Paxos executes as follows. When a client wants to propose a
value $x$, it sends $x$ to a proposer $p$. Upon receiving $x$, $p$ begins
executing one round of Paxos, say round $i$. First, it executes Phase 1. It
sends $\PhaseIA{i}$ messages to at least a majority of the acceptors. An
acceptor ignores a $\PhaseIA{i}$ message if it has already received a message
in a larger round. Otherwise, it replies with a $\PhaseIB{i}{vr}{vv}$ message
containing the largest round $vr$ in which the acceptor voted and the value it
voted for, $vv$. If the acceptor hasn't voted yet, then $vr = -1$ and $vv = $
\textsf{null}. When the proposer receives \textsc{Phase1b} messages from a
majority of the acceptors, Phase 1 ends and Phase 2 begins.

%

At the start of Phase 2, the proposer uses the \textsc{Phase1B} messages that
it received in Phase 1 to select a value $x$ such that no value other than $x$
has been or will be chosen in any round less than $i$. Specifically $x$ is the
vote value associated with the largest received vote round, or any value if no
acceptor had voted (see~\cite{lamport2001paxos} for details). Then, the
proposer sends $\PhaseIIA{i}{x}$ messages to at least a majority of the
acceptors. An acceptor ignores a $\PhaseIIA{i}{x}$ message if it has already
received a message in a larger round. Otherwise, it votes for $x$ and sends
back a $\PhaseIIB{i}$ message to the proposer. If a majority of acceptors vote
for the value (i.e.\ if the proposer receives $\PhaseIIB{i}$ messages from at
least a majority of the acceptors), then the value is chosen, and the proposer
informs the client.  This execution is illustrated in
\figref{PaxosBackgroundDiagram}.

\subsection{Flexible Paxos}
Paxos deploys a set of $2f+1$ acceptors, and proposers communicate with at
least a \emph{majority} of the acceptors in Phase 1 and in Phase 2.
\defword{Flexible Paxos}~\cite{howard2017flexible} is a Paxos variant that
generalizes the notion of a \emph{majority} to that of a \emph{quorum}.
Specifically, Flexible Paxos introduces the notion of a \defword{configuration}
$C = (A; P1; P2)$. $A$ is a set of acceptors. $P1$ and $P2$ are sets of
\defword{quorums}, where each quorum is a subset of $A$. A configuration
satisfies the property that every quorum in $P1$ (known as a \defword{Phase 1
quorum}) intersects every quorum in $P2$ (known as a \defword{Phase 2 quorum}).
%


Flexible Paxos is identical to Paxos with the exception that proposers now
communicate with an arbitrary Phase 1 quorum in Phase 1 and an arbitrary Phase
2 quorum in Phase 2. In the remainder of this paper, we assume that all
protocols operate using quorums from an arbitrary configuration rather than
majorities from a fixed set of $2f+1$ acceptors. So when we say Paxos, for
example, we really mean Flexible Paxos. Note that configurations are also known
as quorum systems~\cite{vukolic2013origin}.

}
{\section{Matchmaker Paxos}\seclabel{MatchmakerPaxos}
We now present \defword{Matchmaker Paxos}, a Paxos variant that allows us to
reconfigure the set of acceptors. After we build some intuition about the
protocol, we describe the protocol in detail and prove it is safe. In the next
section, we'll extend Matchmaker Paxos to Matchmaker MultiPaxos.

\subsection{Overview and Intuition}
Matchmaker Paxos is largely identical to Paxos. Like Paxos, a Matchmaker Paxos
deployment includes an arbitrary number of clients, a set of at least $f+1$
proposers, and some set of acceptors, as illustrated in
\figref{MatchmakerPaxosDiagram}. Paxos assumes that a \emph{single, fixed}
configuration of acceptors is used for every round.  The big difference between
Paxos and Matchmaker Paxos is that Matchmaker Paxos allows every round to have
a \emph{different} configuration of acceptors. Round $0$ may use some
configuration $C_0$, while round $1$ may use some completely different
configuration $C_1$. This idea was first introduced by Vertical
Paxos~\cite{lamport2009vertical}.

{
\tikzstyle{proc}=[draw, circle, thick, inner sep=2pt]
\tikzstyle{client}=[proc, fill=clientcolor!25]
\tikzstyle{proposer}=[proc, fill=proposercolor!25]
\tikzstyle{matchmaker}=[proc, fill=matchmakercolor!25]
\tikzstyle{acceptor0}=[proc, fill=config0color!25]
\tikzstyle{acceptor1}=[proc, fill=config1color!25]

\tikzstyle{proclabel}=[inner sep=0pt, align=center, font=\small]

\tikzstyle{comm}=[-latex, thick]
\tikzstyle{commnum}=[fill=white, text=black, inner sep=0pt]

\begin{figure}[ht]
  \centering
  \begin{tikzpicture}[xscale=1.5]
    \node[client] (c1) at (0, 2) {$c_1$};
    \node[client] (c2) at (0, 1) {$c_2$};
    \node[client] (c3) at (0, 0) {$c_3$};
    \node[proposer] (p1) at (1, 1.5) {$p_1$};
    \node[proposer] (p2) at (1, 0.5) {$p_2$};
    \node[matchmaker] (m1) at (2.00, 3.00) {$m_1$};
    \node[matchmaker] (m2) at (2.50, 2.75) {$m_2$};
    \node[matchmaker] (m3) at (3.00, 2.50) {$m_3$};
    \node[acceptor0] (a1) at (3, 1.75) {$a_1$};
    \node[acceptor0] (a2) at (3, 1.00) {$a_2$};
    \node[acceptor0] (a3) at (3, 0.25) {$a_3$};
    \node[acceptor1] (b1) at (2.00, -1.00) {$b_1$};
    \node[acceptor1] (b2) at (2.50, -0.75) {$b_2$};
    \node[acceptor1] (b3) at (3.00, -0.50) {$b_3$};

    \node[proclabel] (clients) at (0, 4) {Clients};
    \node[proclabel] (proposers) at (1, 4) {$f+1$\\Proposers};
    \node[proclabel] (matchmakers) at (2.5, 4) {$2f+1$\\Matchmakers};
    \node[proclabel] (oldacceptors) at (4, 1) {$C_0$ Acceptors};
    \node[proclabel] (newacceptors) at (2.5, -1.75) {$C_1$ Acceptors};
    \halffill{clients}{clientcolor!25}
    \quarterfill{proposers}{proposercolor!25}
    \quarterfill{matchmakers}{matchmakercolor!25}
    \halffill{oldacceptors}{config0color!25}
    \halffill{newacceptors}{config1color!25}

    \draw[comm, bend left] (c1) to node[commnum]{1} (p1);
    \draw[comm, bend left] (p1) to node[commnum]{8} (c1);

    \draw[comm, matchmakercolor, bend left=45] (p1) to node[commnum]{2} (m1);
    \draw[comm, matchmakercolor, bend left=25] (p1) to node[commnum]{2} (m2);
    \draw[comm, matchmakercolor, bend right=30] (m1) to node[commnum]{3} (p1);
    \draw[comm, matchmakercolor, bend right=10] (m2) to node[commnum]{3} (p1);

    \draw[comm, config0color, bend left=20] (p1) to node[commnum]{4} (a1);
    \draw[comm, config0color, bend left=5] (p1) to node[commnum]{4} (a2);
    \draw[comm, config0color, bend left=0] (a1) to node[commnum]{5} (p1);
    \draw[comm, config0color, bend left=15] (a2) to node[commnum]{5} (p1);

    \draw[comm, config1color, bend left=10] (p1) to node[commnum]{6} (b1);
    \draw[comm, config1color, bend left=20] (p1) to node[commnum]{6} (b2);
    \draw[comm, config1color, bend left=0] (b1) to node[commnum]{7} (p1);
    \draw[comm, config1color, bend right=8] (b2) to node[commnum]{7} (p1);

    \node[draw=gray!50, thick, align=center] at (2, -2.75) {
      \tikz[baseline=-0.75ex]{\draw[comm, matchmakercolor] (0, 0) to node[pos=0.45, commnum]{2/3} (1.5, 0);} Matchmaking Phase \\[4pt]
      \tikz[baseline=-0.75ex]{\draw[comm, config0color] (0, 0) to node[pos=0.45, commnum]{4/5} (1.5, 0);} Phase 1 \qquad
      \tikz[baseline=-0.75ex]{\draw[comm, config1color] (0, 0) to node[pos=0.45, commnum]{6/7} (1.5, 0);} Phase 2
    };
  \end{tikzpicture}%
  \caption{Matchmaker Paxos communication diagram.}%
  \figlabel{MatchmakerPaxosDiagram}
\end{figure}
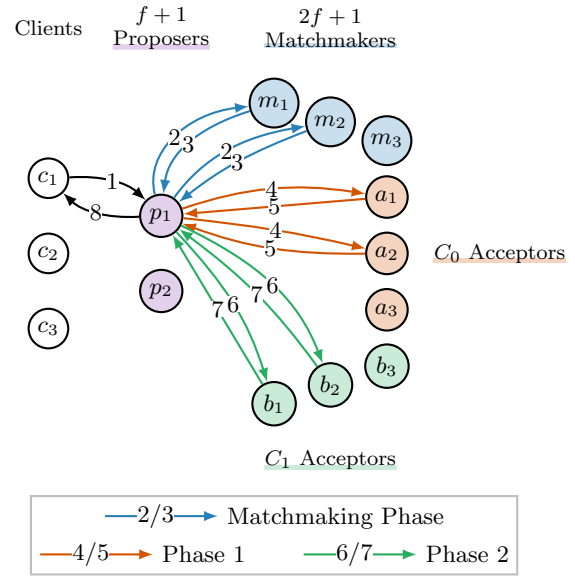}

Recall from \secref{Background}, that a Paxos proposer in round $i$ executes
Phase $1$ in order to (1) learn of any value that may have been chosen in a
round less than $i$ and (2) prevent any new values from being chosen in any
round less than $i$. To do so, the proposer contacts the \emph{fixed set} of
acceptors. A Matchmaker Paxos proposer is no different. It must also execute
Phase $1$ and establish that these two properties hold. The difference is that
there is no longer a single fixed configuration of acceptors to contact.
Instead, a Matchmaker Paxos proposer has to contact all of the configurations
used in rounds less than $i$.

However, every round uses a different configuration of acceptors, so how does
the proposer of round $i$ know which acceptors to contact in Phase 1? To
resolve this question, a Matchmaker Paxos deployment also includes a set of
$2f+1$ \defword{matchmakers}. When a proposer begins executing round $i$, it
selects a configuration $C_i$. It sends the configuration $C_i$ to the
matchmakers, and the matchmakers reply with the configurations used in previous
rounds. We call this the Matchmaking phase. The proposer then executes Phase 1
of Paxos with these prior configurations, and then executes Phase 2 with
configuration $C_i$, as illustrated in \figref{MatchmakerPaxosDiagram}.
Naively, the extra round trip of communication with the matchmakers and the
large number of configurations in Phase 1 make Matchmaker Paxos slow. In
Sections \ref{sec:Optimizations}, \ref{sec:MultiOptimization}, and
\ref{sec:RetiringOldConfigurations}, we'll introduce optimizations to eliminate
these costs.
\begin{revisions}
  It may also seem like we've solved the problem of reconfiguring acceptors by
  introducing a new problem of reconfiguring matchmakers, but this is not the
  case. In \secref{ReconfiguringMatchmakers}, we explain how we can reconfigure
  the matchmakers in a straightforward way.
\end{revisions}

\subsection{Details}
Every matchmaker maintains a log $L$ of configurations indexed by round. That
is, $L[i]$ stores the configuration of round $i$. When a proposer receives a
request $x$ from a client and begins executing round $i$, it first selects a
configuration $C_i$ to use in round $i$. It then sends a $\MatchA{i}{C_i}$
message to all of the matchmakers.

When a matchmaker receives a $\MatchA{i}{C_i}$ message, it checks to see if it
had previously received a $\MatchA{j}{C_j}$ message for some round $j \geq i$.
If so, the matchmaker ignores the $\MatchA{i}{C_i}$ message. Otherwise, it
inserts $C_i$ in log entry $i$ and computes the set $H_i$ of previous
configurations in the log: $H_i = \setst{(j, C_j)}{j < i, C_j \in L}$. It then
replies to the proposer with a $\MatchB{i}{H_i}$ message.
An example execution of a matchmaker is illustrated in \figref{MatchmakerLogs}.
Matchmaker pseudocode is given in \algoref{MatchmakerPseudocode}.

{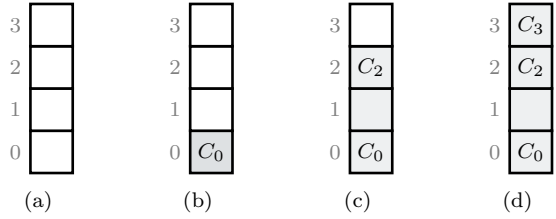
\begin{figure}[ht]
  \tikzstyle{logentry}=[draw, line width=1pt, minimum width=16pt,
                        minimum height=16pt]
  \newcommand{\mlog}[9]{
    \node[logentry, fill=#6, label={[gray]180:0}] (l0) at (0, 0) {#1};
    \node[logentry, fill=#7, label={[gray]180:1}, above=-1pt of l0] (l1) {#2};
    \node[logentry, fill=#8, label={[gray]180:2}, above=-1pt of l1] (l2) {#3};
    \node[logentry, fill=#9, label={[gray]180:3}, above=-1pt of l2] (l3) {#4};
  }

  \small
  \begin{subfigure}[b]{0.2\columnwidth}
    \centering
    \begin{tikzpicture}
      \mlog{}{}{}{}{}{white}{white}{white}{white}
    \end{tikzpicture}
    \caption{}
    \figlabel{MatchmakersA}
  \end{subfigure}\hspace{0.04\columnwidth}
  \begin{subfigure}[b]{0.2\columnwidth}
    \centering
    \begin{tikzpicture}
      \mlog{$C_0$}{}{}{}{}{flatgray!50}{white}{white}{white}
    \end{tikzpicture}
    \caption{}\figlabel{MatchmakersB}
  \end{subfigure}\hspace{0.04\columnwidth}
  \begin{subfigure}[b]{0.2\columnwidth}
    \centering
    \begin{tikzpicture}
      \mlog{$C_0$}{}{$C_2$}{}{}{flatgray!25}{flatgray!25}{flatgray!25}{white}
    \end{tikzpicture}
    \caption{}\figlabel{MatchmakersC}
  \end{subfigure}\hspace{0.04\columnwidth}
  \begin{subfigure}[b]{0.2\columnwidth}
    \centering
    \begin{tikzpicture}
      \mlog{$C_0$}{}{$C_2$}{$C_3$}{}%
           {flatgray!25}{flatgray!25}{flatgray!25}{flatgray!25}
    \end{tikzpicture}
    \caption{}\figlabel{MatchmakersD}
  \end{subfigure}

  \caption{%
    A matchmaker's log over time.
    (a) Initially, the matchmaker's log is empty.
    (b) Then, the matchmaker receives $\MatchA{0}{C_0}$. It inserts
    $C_0$ in log entry $0$ and returns $\MatchB{0}{\emptyset}$ since
    the log does not contain any configuration in any round less than $0$.
    (c) The matchmaker then receives $\MatchA{2}{C_2}$. It inserts $C_2$
    in log entry $2$ and returns $\MatchB{2}{\set{(0, C_0)}}$.
    (d) It then receives $\MatchA{3}{C_3}$, inserts
    $C_3$ in log entry $3$, and returns $\MatchB{3}{\set{(0, C_0),
    (2, C_2)}}$.
    At this point, if the matchmaker were to receive $\MatchA{1}{C_1}$, it
    would ignore it.
  }\figlabel{MatchmakerLogs}
\end{figure}
}
{\begin{algorithm}
  \begin{algorithmic}[1]
    \GlobalState a log $L$ indexed by round, initially empty

    \Upon{receiving $\MatchA{i}{C_i}$ from proposer $p$}
      \If{$\exists$ a configuration $C_j$ in round $j \geq i$ in $L$}
      \State ignore the $\MatchA{i}{C_i}$ message
      \Else
        \State $H_i \gets \setst{(j, C_j)}{C_j \in L}$
        \State $L[i] \gets C_i$
        \State send $\MatchB{i}{H_i}$ to $p$
      \EndIf
    \EndUpon
  \end{algorithmic}
  \caption{Matchmaker Pseudocode}\algolabel{MatchmakerPseudocode}
\end{algorithm}

}

When the proposer in round $i$ receives $\MatchB{i}{H_i^1}$, $\ldots$,
$\MatchB{i}{H_i^{f+1}}$ from $f+1$ matchmakers, it computes $H_i =
\cup_{j=1}^{f+1} H_i^j$. For example, with $f=1$ and $i=2$, if the proposer in
round $2$ receives $\MatchB{2}{\set{(0, C_0)}}$ and $\MatchB{2}{\set{(1,
C_1)}}$, it computes $H_2 = \set{(0, C_0), (1, C_1)}$. Note that every round is
statically assigned to a single proposer and that a proposer selects a single
configuration for a round. So, if two matchmakers return configurations for the
same round, they are guaranteed to be the same.

The proposer then ends the Matchmaking phase and begins Phase 1. It sends
$\textsc{Phase1A}$ messages to every acceptor in every configuration in $H_i$
and waits to receive $\textsc{Phase1B}$ messages from at least a Phase 1 quorum
from every configuration in $H_i$. Using the previous example, the proposer
would send $\textsc{Phase1A}$ messages to every acceptor in $C_0$ and $C_1$ and
would wait for $\textsc{Phase1B}$ messages from a Phase 1 quorum of $C_0$ and a
Phase 1 quorum of $C_1$. The proposer then runs Phase 2 with $C_i$.

Acceptor and proposer pseudocode are shown in \algoref{AcceptorPseudocode} and
\algoref{ProposerPseudocode} respectively. To keep things simple, we assume
that round numbers are integers, but generalizing is straightforward. A
Matchmaker Paxos acceptor is identical to a Paxos acceptor. A Matchmaker Paxos
proposer is nearly identical to a regular Paxos proposer with the exception of
the Matchmaking phase and the configurations used in Phase 1 and Phase 2.  For
clarity of exposition, we omit straightforward details surrounding re-sending
dropped messages and nacking ignored messages.

{\begin{algorithm}[t]
  \begin{algorithmic}[1]
    \GlobalState the largest seen round $r$, initially $-1$
    \GlobalState the largest round $vr$ voted in, initially $-1$
    \GlobalState the value $vv$ voted for in round $vr$, initially \textsf{null}

    \Upon{receiving $\PhaseIA{i}$ from $p$ with $i > r$}
      \State $r \gets i$
      \State send $\PhaseIB{i}{vr}{vv}$ to $p$
    \EndUpon

    \Upon{receiving $\PhaseIIA{i}{x}$ from $p$ with $i \geq r$}
      \State $r, vr, vv \gets i, i, x$
      \State send $\PhaseIIB{i}$ to $p$
    \EndUpon
  \end{algorithmic}
  \caption{Acceptor Pseudocode}\algolabel{AcceptorPseudocode}
\end{algorithm}
}
{\begin{algorithm}[t]
  \begin{algorithmic}[1]
    \GlobalState a value $x$, initially \textsf{null}
    \GlobalState a round $i$, initially $-1$
    \GlobalState the configuration $C_i$ for round $i$, initially \textsf{null}
    \GlobalState the prior configurations $H_i$ for round $i$, initially \textsf{null}

    \Upon{receiving value $y$ from a client}
      \State $i \gets$ next largest round owned by this proposer
      \State $x \gets y$
      \State $C_i \gets$ an arbitrary configuration
      \State send $\MatchA{i}{C_i}$ to all of the matchmakers
    \EndUpon

    \Upon{%
      receiving $\MatchB{i}{H_i^1}, \ldots, \MatchB{i}{H_i^{f+1}}$ from $f+1$
      matchmakers
    }
      \State $H_i \gets \bigcup_{j=1}^{f+1} H_i^j$
      \State send $\PhaseIA{i}$ to every acceptor in $H_i$
    \EndUpon

    \Upon{%
      receiving $\PhaseIB{i}{-}{-}$ from a Phase 1 quorum from every
      configuration in $H_i$
    }
      \State $k \gets$ the largest $vr$ in any $\PhaseIB{i}{vr}{vv}$
      \If{$k \neq -1$}
        \State \markrevisions{$x \gets$ the corresponding $vv$ in round $k$}
      \EndIf
      \State \markrevisions{send $\PhaseIIA{i}{x}$ to every acceptor in $C_i$}
    \EndUpon

    \Upon{receiving $\PhaseIIB{i}$ from a Phase 2 quorum}
      \State $x$ is chosen, inform the client
    \EndUpon
  \end{algorithmic}
  \caption{Proposer Pseudocode}\algolabel{ProposerPseudocode}
\end{algorithm}
}

\subsection{Proof of Safety}
We now prove that Matchmaker Paxos is safe; i.e.\ every execution of Matchmaker
Paxos chooses at most one value. We include the proof not only to convince you
that the protocol is safe, but also because we believe that the proof makes it
clear how and why Matchmaker Paxos works the way it does.

\begin{proof}
Our proof is based off of the Paxos safety proof in~\cite{lamport2006fast}. We
prove, for every round $i$, the statement $P(i)$: ``if a proposer proposes a
value $v$ in round $i$ (i.e.\ sends a \textsc{Phase2A} message for value $v$ in
round $i$), then no value other than $v$ has been or will be chosen in any
round less than $i$.'' $P(i)$ suffices to prove that Matchmaker Paxos is safe
for the following reason. Assume for contradiction that Matchmaker Paxos
chooses distinct values $x$ and $y$ in rounds $i$ and $j$ with $i < j$.  Some
proposer must have proposed $y$ in round $j$, so $P(j)$ ensures us that no
value other than $y$ could have been chosen in round $i$. But, $x$ was chosen,
a contradiction.

%

We prove $P(i)$ by strong induction on $i$. $P(0)$ is vacuous because there are
no rounds less than $0$. For the general case $P(i)$, we assume $P(0), \ldots,
P(i-1)$. We perform a case analysis on the proposer's pseudocode. Either $k$ is
$-1$ or it is not (line 11). First, assume it is not. In this case, the
proposer proposes $x$, the value proposed in round $k$ (line 12). We perform a
case analysis on $j$ to show that no value other than $x$ has been or will be
chosen in any round $j < i$.


\textbf{Case 1: $j > k$.}
We show that no value has been or will be chosen in round $j$.
Recall that at the end of the Matchmaking phase, the proposer computed the
set $H_i$ of prior configurations using responses from a set $M$ of $f+1$
matchmakers. Either $H_i$ contains a configuration $C_j$ in round $j$ or it
doesn't.

First, suppose it does. Then, the proposer sent $\PhaseIA{i}$ messages to all
of the acceptors in $C_j$. A Phase 1 quorum of these acceptors, say $Q$, all
received $\PhaseIA{i}$ messages and replied with \textsc{Phase1B} messages.
Thus, every acceptor in $Q$ set its round $r$ to $i$, and in doing so, promised
to never vote in any round less than $i$.  Moreover, none of the acceptors in
$Q$ had voted in any round greater than $k$. So, every acceptor in $Q$ has not
voted and never will vote in round $j$. For a value $v'$ to be chosen in round
$j$, it must receive votes from some Phase 2 quorum $Q'$ of round $j$
acceptors. But, $Q$ and $Q'$ necessarily intersect, so this is impossible.
Thus, no value has been or will be chosen in round $j$.

Now suppose that $H_i$ does \emph{not} contain a configuration for round $j$.
$H_i$ is the union of $f+1$ $\textsc{MatchB}$ messages from the $f+1$
matchmakers in $M$. Thus, if $H_i$ does not contain a configuration for round
$j$, then none of the $\textsc{MatchB}$ messages did either. This means that
for every matchmaker $m \in M$, when $m$ received $\MatchA{i}{C_i}$, it did not
contain a configuration for round $j$ in its log. Moreover, by processing the
$\MatchA{i}{C_i}$ request and inserting $C_i$ in log entry $i$, the matchmaker
is guaranteed to never process a $\MatchA{j}{C_j}$ request in the future. Thus,
every matchmaker in $M$ has not processed a \textsc{MatchA} request in round
$j$ and never will. For a value to be chosen in round $j$, the proposer
executing round $j$ must first receive replies from $f+1$ matchmakers, say
$M'$, in round $j$. But, $M$ and $M'$ necessarily intersect, so this is
impossible. Thus, no value has been or will be chosen in round $j$.

\textbf{Case 2: $j = k$.}
In a given round, at most one value is proposed, let alone chosen. $x$ is
\emph{the} value proposed in round $k$, so no other value could be chosen in
round $k$.

\textbf{Case 3: $j < k$.}
\begin{revisions}
  We can apply the inductive hypothesis to get $P(k)$ which states that no
  value other than $x$ has been or will be chosen in any round less than $k$.
  This includes round $j$, which is exactly what we're trying to prove.
\end{revisions}

Finally, if $k$ is $-1$, then we are in the same situation as in Case 1. No
value has been or will be chosen in any round less than $i$.
\end{proof}


\subsection{Optimizations}\seclabel{Optimizations}

\textbf{Optimization 1: Proactive Matchmaking.}
Note that the Matchmaking phase is independent of any client value, so a
proposer can proactively run the Matchmaking phase in round $i$ \emph{before}
it hears from a client. This is similar to proactively executing Phase 1, a
standard optimization.


\textbf{Optimization 2: Phase 1 Bypassing.}
Assume that the proposer in round $i$ has proactively executed the Matchmaking
phase and Phase 1. Through Phase 1, it finds that $k=-1$ and thus learns that
no value has been chosen in any round less than $i$ (see the safety proof
above). Assume that before executing Phase 2, the proposer transitions from
round $i$ to round $i+1$, which it also owns. After executing the Matchmaking
phase in round $i+1$, the proposer can skip Phase 1 and proceed directly to
Phase 2. Why? The proposer established in round $i$ that no value has been or
will be chosen in any round less than $i$. Moreover, because it did not run
Phase $2$ in round $i$, it also knows that no value has been or will be chosen
in round $i$. Together, these imply that no value has been or will be chosen in
any round less than $i+1$. Normally, the proposer would run Phase $1$ in round
$i+1$ to establish this fact, but since it has already established it, it can
instead proceed directly to Phase 2.

\begin{techreport}

Similarly, assume that the proposer in round $i$ has executed the Matchmaking
phase and Phase 1. It then proposes a value $v$ in round $i$. The proposer then
transitions from round $i$ to round $i+1$, which it also owns. After executing
the Matchmaking phase in round $i+1$, the proposer can skip Phase 1 and proceed
directly to Phase 2 to propose $v$. Why? The proposer established in round $i$
that no value other than $v$ has been or will be chosen in any round less than
$i$. Moreover, no value other than $v$ has been or will be chosen in round $i$.
Together, these imply that no value other than $v$ has been or will be chosen
in any round less than $i+1$. Normally, the proposer would run Phase $1$ in
round $i+1$ to establish this fact, but since it has already established it, it
can instead proceed immediately to Phase 2.
\end{techreport}

\newcommand{\red}[1]{\textcolor{flatred}{#1}}
\newcommand{\blue}[1]{\textcolor{flatblue}{#1}}
This optimization depends on a proposer being the leader of round $i$
\emph{and} the leader of the next round $i+1$. We can construct a set of rounds
such that this is always the case \markrevisions{(remember that the set of
rounds can be \emph{any} unbounded totally ordered set with a least element)}.
Let the set of rounds be the set of lexicographically ordered tuples $(r, id,
s)$ where $r$ and $s$ are both integers and $id$ is a proposer id. A proposer
owns all the rounds that contain its id. For example given two proposers
$\red{a}$ and $\blue{b}$, we have the following ordering on rounds:
\begin{gather*}
  (0, \red{a}, 0) < (0, \red{a}, 1) < (0, \red{a}, 2) < (0, \red{a}, 3) < \cdots \\
  (0, \blue{b}, 0) < (0, \blue{b}, 1) < (0, \blue{b}, 2) < (0, \blue{b}, 3) < \cdots \\
  (1, \red{a}, 0) < (1, \red{a}, 1) < (1, \red{a}, 2) < (1, \red{a}, 3) < \cdots
\end{gather*}
With this set of rounds, the proposer $p$ in round $(r, p, s)$ always owns the
next round $(r, p, s + 1)$.

We acknowledge that, for now, this optimization sounds very esoteric. For
Matchmaker Paxos, it is. In the next section, we'll see that this optimization
is essential for implementing Matchmaker MultiPaxos with good performance.

\begin{techreport}

  Also note that this optimization is not particular to Matchmaker Paxos. Paxos
  and MultiPaxos can both take advantage of this optimization.
\end{techreport}

\textbf{Optimization 3: Garbage Collection.}
A proposer performs Phase 1 with every configuration returned by the
matchmakers. This can be prohibitively expensive if the matchmakers return a
large number of configurations. Later in \secref{RetiringOldConfigurations}, we
introduce a garbage collection protocol to delete old configurations. In
\secref{Evaluation}, we see empirically that Matchmakers usually return just a
single configuration.

\begin{techreport}

\textbf{Optimization 4: Round Pruning.}
Note that a proposer does not always need to receive \textsc{Phase1B} messages
from \emph{all} previous rounds. If the proposer in round $i$ receives a
$\PhaseIB{i}{vr}{vv}$ message, then it no longer needs to intersect
configurations in rounds less than $vr$. This optimization is a lighter weight
version of the garbage collection protocol we describe in
\secref{RetiringOldConfigurations}.

\textbf{Optimization 5: Concurrent Matchmaking, Phase 1.}
Note that a proposer can safely execute the Matchmaking phase and Phase 1 in
parallel. This is useful if proposers rarely change the set of acceptors during
a leader change.  It is particularly useful if we co-locate the $2f+1$
matchmakers with $2f+1$ acceptors and combine the Matchmaking and Phase 1
messages together. This makes Matchmaker Paxos have equal message complexity to
MultiPaxos' horizontal reconfiguration protocol.

\textbf{Optimization 6: Flexible Matchmakers.}
For simplicity, we described matchmakers as a fixed set of $2f+1$ matchmakers.
In reality, we can deploy an arbitrary set of matchmakers with an arbitrary set
of matchmaker quorums, so long as every pair of quorums intersect.
\end{techreport}

%
%
%
}
{\section{Matchmaker MultiPaxos}\seclabel{MatchmakerMultiPaxos}
In this section, we extend Matchmaker Paxos to Matchmaker MultiPaxos.

\subsection{MultiPaxos}
First, we summarize MultiPaxos. Whereas Paxos is a consensus protocol that
agrees on a single value, \defword{MultiPaxos}~\cite{lamport1998part,
van2015paxos} is a \defword{state machine replication protocol} that agrees on
a sequence, or ``log'' of values. MultiPaxos manages multiple
\defword{replicas} of a state machine.  Clients send state machine commands to
MultiPaxos, MultiPaxos places the commands in a totally ordered log, and state
machine replicas execute the commands in log order. By beginning in the same
initial state and executing the same commands in the same order, all state
machine replicas are kept in sync.
%
%
%

{
\tikzstyle{proc}=[draw, circle, thick, inner sep=2pt]
\tikzstyle{client}=[proc, fill=clientcolor!25]
\tikzstyle{proposer}=[proc, fill=proposercolor!25]
\tikzstyle{acceptor}=[proc, fill=acceptorcolor!25]
\tikzstyle{replica}=[proc, fill=replicacolor!25]

\tikzstyle{proclabel}=[inner sep=0pt, align=center]

\tikzstyle{comm}=[-latex, thick]
\tikzstyle{commnum}=[fill=white, inner sep=0pt]

\begin{figure}[h]
  \centering
  \begin{tikzpicture}[xscale=2]
    \node[client] (c1) at (0, 2) {$c_1$};
    \node[client] (c2) at (0, 1) {$c_2$};
    \node[client] (c3) at (0, 0) {$c_3$};
    \node[proposer] (p1) at (1, 1.5) {$p_1$};
    \node[proposer] (p2) at (1, 0.5) {$p_2$};
    \node[acceptor] (a1) at (2, 2) {$a_1$};
    \node[acceptor] (a2) at (2, 1) {$a_2$};
    \node[acceptor] (a3) at (2, 0) {$a_3$};
    \node[replica] (r1) at (3, 1.5) {$r_1$};
    \node[replica] (r2) at (3, 0.5) {$r_2$};

    \crown{(p1.north)++(0,-0.15)}{0.25}{0.25}
    \node[proclabel] (clients) at (0, 3) {Clients};
    \node[proclabel] (proposers) at (1, 3) {$f+1$\\Proposers};
    \node[proclabel] (acceptors) at (2, 3) {Configuration $C$\\of Acceptors};
    \node[proclabel] (replicas) at (3, 3) {$f+1$\\Replicas};
    \halffill{clients}{clientcolor!25}
    \quarterfill{proposers}{proposercolor!25}
    \quarterfill{acceptors}{acceptorcolor!25}
    \quarterfill{replicas}{replicacolor!25}

    \draw[comm] (c1) to node[commnum]{1} (p1);
    \draw[comm, bend left=15] (p1) to node[commnum]{2} (a1);
    \draw[comm, bend left=15] (p1) to node[commnum]{2} (a2);
    \draw[comm, bend left=15] (a1) to node[commnum]{3} (p1);
    \draw[comm, bend left=15] (a2) to node[commnum]{3} (p1);
    \draw[comm] (p1) to node[commnum]{4} (r1);
    \draw[comm, bend right=40] (p1) to node[commnum]{4} (r2);
    \draw[comm, bend right=45] (r1) to node[commnum]{5} (c1);
  \end{tikzpicture}
  \caption{%
    An example execution of MultiPaxos ($f=1$). The leader is adorned with a
    crown.
  }%
  \figlabel{MultiPaxosBackgroundDiagram}
\end{figure}
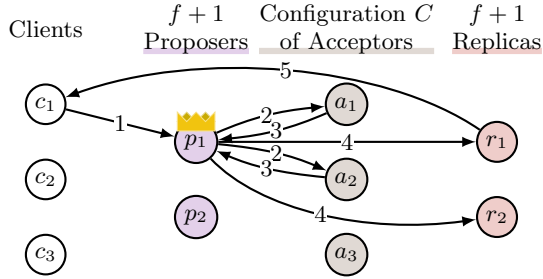}

To agree on a log of commands, MultiPaxos implements one instance of Paxos for
every log entry. The $i$th instance of Paxos chooses the command in
log entry $i$. More concretely, a MultiPaxos deployment that tolerates $f$
faults consists of an arbitrary number of clients, at least $f+1$ proposers, a
configuration $C$ of acceptors, and at least $f+1$ replicas, as illustrated in
\figref{MultiPaxosBackgroundDiagram}.

One of the proposers is elected leader in some round, say round $i$. We assume
the leader knows that log entries up to and including log entry $w$ have
already been chosen (e.g., by communicating with previous leaders, or by
communicating with the replicas). The leader then runs Phase 1 of Paxos in
round $i$ for \emph{every} log entry larger than $w$. Note that even though
there are an infinite number of log entries larger than $w$, the leader can
execute Phase 1 using a finite amount of information.  In particular, the
leader sends a single $\PhaseIA{i}$ message that acts as the \textsc{Phase1A}
message for every log entry larger than $w$. Also, an acceptor replies with a
$\PhaseIB{i}{vr}{vv}$ message only for log entries in which the acceptor has
voted. The infinitely many log entries in which the acceptor has not yet voted
do not yield an explicit \textsc{Phase1B} message.

%

The leader's knowledge about the log after Phase 1 is shown in
\figref{MultiPaxosPhase1Log}. The leader knows that a prefix of the log (up to
and including log entry $w$) has already been chosen. Then, there is a
subsequence of the log that contains commands that may have already been chosen
in a round less than $i$. This subsequence may contain unchosen entries, which
we call ``holes''. Finally, the log has an infinite tail of empty entries in
which the leader knows no command has previously been chosen.

{\begin{figure}[ht]
  \centering
  \tikzstyle{logentry}=[draw, line width=1pt, minimum width=18pt,
                        minimum height=18pt]
  \tikzstyle{lognum}=[gray, font=\scriptsize]
  \tikzstyle{chosen}=[fill=flatred!15]
  \tikzstyle{maybe}=[fill=flatyellow!15]
  \tikzstyle{notchosen}=[]
  \tikzstyle{brace}=[decorate, decoration={brace, mirror, amplitude=4pt},
                     line width=1pt]
  \tikzstyle{chosenlabel}=[align=center, below=6pt]
  \begin{tikzpicture}
    \node[logentry, label={[lognum]0}, chosen] (l0) at (0, 0) {$a$};
    \node[logentry, label={[lognum]1}, chosen, right=-1pt of l0] (l1) {$b$};
    \node[logentry, label={\scriptsize $w$}, chosen, right=-1pt of l1] (l2) {$c$};
    \node[logentry, label={[lognum]3}, maybe, right=-1pt of l2] (l3) {$d$?};
    \node[logentry, label={[lognum]4}, maybe, right=-1pt of l3] (l4) {};
    \node[logentry, label={[lognum]5}, maybe, right=-1pt of l4] (l5) {$e$?};
    \node[logentry, label={[lognum]6}, notchosen, right=-1pt of l5] (l6) {};
    \node[logentry, label={[lognum]7}, notchosen, right=-1pt of l6] (l7) {};
    \node[logentry, label={[lognum]8}, notchosen, right=-1pt of l7] (l8) {};
    \node[logentry, draw=white, right=0 of l8] (nodes) {$\cdots$};

    \draw[brace] ([xshift=1pt] l0.south west) to %
                 node[chosenlabel]{already\\chosen} %
                 ([xshift=-1pt] l2.south east);
    \draw[brace] ([xshift=1pt] l3.south west) to %
                 node[chosenlabel]{maybe\\chosen} %
                 ([xshift=-1pt] l5.south east);
    \draw[brace] ([xshift=1pt] l6.south west) to %
                 node[chosenlabel]{not\\chosen} %
                 ([xshift=-1pt] nodes.south east);
  \end{tikzpicture}
  \caption{A leader's knowledge of the log after Phase 1.}%
  \figlabel{MultiPaxosPhase1Log}
\end{figure}
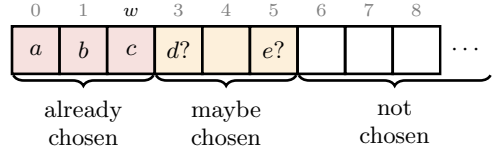}

After Phase 1, the leader sends a \textsc{Phase2A} message for every log entry
in the middle subsequence of potentially chosen commands, proposing a ``no-op''
command for the holes. Simultaneously, the leader begins accepting client
requests. When a client wants to propose a state machine command, it sends the
command to the leader. The leader assigns log entries to commands in increasing
order, beginning at the start of the infinite tail of unchosen entries. It then
runs Phase 2 of Paxos to get the command chosen in that entry in round $i$.
Once the leader learns that a command has been chosen in a given log entry, it
informs the replicas. Replicas insert chosen commands into their logs and
execute the logs in prefix order, sending the results of execution back to the
clients. This execution is illustrated in \figref{MultiPaxosBackgroundDiagram}.

%

It is critical to note that a leader performs Phase 1 of Paxos only once
\emph{per round}, not once \emph{per command}. In other words, Phase 1 is not
performed during normal operation. It is \markrevisions{performed only} when
the leader fails and new leader is elected in a larger round, an uncommon
occurrence.

\subsection{Matchmaker MultiPaxos}
We first extend Matchmaker Paxos to Matchmaker MultiPaxos with Optimization 1
(Proactive Matchmaking) but without Optimization 2 (Phase 1 Bypassing)
or 3 (Garbage Collection).  We'll see how to incorporate Optimization 2
momentarily and Optimization 3 in the next section. The extension from
Matchmaker Paxos to Matchmaker MultiPaxos is completely analogous to the
extension of Paxos to MultiPaxos. Matchmaker MultiPaxos reaches consensus on a
totally ordered log of state machine commands, one log entry at a time, using
one instance of Matchmaker Paxos for every log entry.

More concretely, a Matchmaker MultiPaxos deployment consists of an arbitrary
number of clients, at least $f+1$ proposers, a set of $2f+1$ matchmakers, a
dynamic set of acceptors (one configuration per round), and a set of at least
$f+1$ state machine replicas.
We assume, as is standard, that a leader election algorithm is used to select
one of the proposers as a stable leader in some round, say round $i$. The
leader selects a configuration $C_i$ of acceptors that it will use for
\emph{every} log entry. The mechanism by which the configuration is chosen is
an orthogonal concern.

The leader then executes the Matchmaking phase in exactly the same way as in
Matchmaker Paxos (i.e.\ it sends $\MatchA{i}{C_i}$ messages to the matchmakers
and awaits $\MatchB{i}{H_i}$ responses). After the Matchmaking phase completes,
the leader executes Phase 1 for \emph{every} log entry. This is identical to
MultiPaxos, except that the leader uses the configurations returned by the
matchmakers rather than assuming a fixed configuration. Note that Optimization
1 allows the leader to execute the Matchmaking phase and Phase 1 before
receiving any client requests.

The leader then enters Phase 2 and operates exactly as it would in MultiPaxos.
It executes Phase 2 with $C_i$ for any potentially unchosen values it learned
about in Phase 1, and it fills in any holes in the log with no-ops. Moreover,
when it receives a state machine command from a client, it assigns the command
a log entry, runs Phase 2 with the acceptors in $C_i$, and informs the replicas
when the command is chosen. Replicas execute commands in log order and send the
results of executing commands back to the clients.

\subsection{Discussion}
Note that unlike regular MultiPaxos, Matchmaker MultiPaxos does not require a
separate mechanism to perform a reconfiguration. Reconfiguration is baked into
the protocol. To change from some configuration $C_\text{old}$ in round $i$ to
some new configuration $C_\text{new}$, the leader of round $i$ simply advances
to round $i+1$ and selects the new configuration $C_\text{new}$. The new
configuration is active immediately after the Matchmaking phase, a one round
trip delay.
%
%
Though the new configuration is active immediately, it is not safe to
deactivate the acceptors in the old configuration immediately. They are still
needed. In \secref{RetiringOldConfigurations}, we'll see why this is the case
and when it is safe to retire old configurations.

Also note that Matchmaker MultiPaxos does \emph{not} perform the Matchmaking
Phase or Phase $1$ on the critical path of normal execution. Similar to how
MultiPaxos \markrevisions{executes Phase 1 only once} per leader change (and
not once per command), Matchmaker MultiPaxos runs the Matchmaking phase and
Phase 1 only \markrevisions{when a new leader is elected or when a leader
changes its round (e.g., when a leader transitions from round $i$ to round $i +
1$ as part of a reconfiguration)}. In the normal case (i.e.\ during Phase 2),
Matchmaker MultiPaxos and MultiPaxos are identical. Matchmaker MultiPaxos does
not introduce \emph{any} overheads in the normal case.

Further note that configurations do not have to be unique across rounds. The
leader in round $i$ is free to re-use a configuration $C_j$ that was used in
some round $j < i$.

\begin{revisions}
  Finally note that, in theory, the Matchmaking phase of Matchmaker Paxos is
  not fully live (which is fundamentally
  unavoidable~\cite{fischer1985impossibility}). If multiple proposers
  simultaneously execute the Matchmaking phase, it is possible to experience
  ``dueling proposers''~\cite{lamport2001paxos}. Matchmaker MultiPaxos avoids
  this in practice using leader election. A proposer only initiates the
  Matchmaking phase if it is elected leader.
\end{revisions}

\subsection{Optimization}\seclabel{MultiOptimization}
Ideally, Matchmaker MultiPaxos' performance would be unaffected by a
reconfiguration. The latency of every client request and the protocol's overall
throughput would remain constant throughout a reconfiguration.  Matchmaker
MultiPaxos as we've described it so far, however, does not meet this ideal.
During a reconfiguration, a leader must temporarily stop processing client
commands and wait for the reconfiguration to finish before resuming normal
operation.

{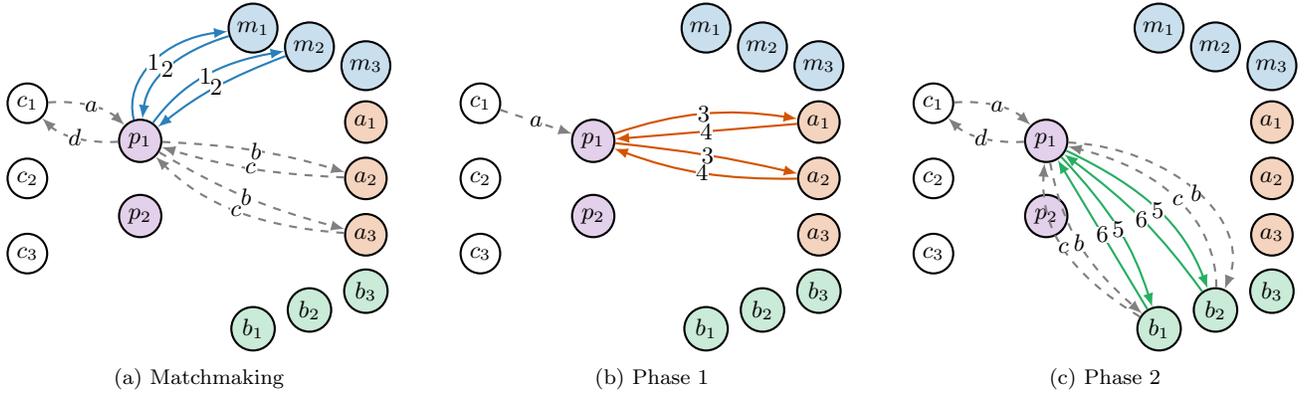
\begin{figure*}[ht]
  \centering

  \tikzstyle{proc}=[draw, circle, thick, inner sep=2pt]
  \tikzstyle{client}=[proc, fill=clientcolor!25]
  \tikzstyle{proposer}=[proc, fill=proposercolor!25]
  \tikzstyle{matchmaker}=[proc, fill=matchmakercolor!25]
  \tikzstyle{acceptor0}=[proc, fill=config0color!25]
  \tikzstyle{acceptor1}=[proc, fill=config1color!25]

  \tikzstyle{proclabel}=[inner sep=0pt, align=center, font=\small]

  \tikzstyle{comm}=[-latex, thick]
  \tikzstyle{clientcomm}=[comm, dashed, gray]
  \tikzstyle{commnum}=[fill=white, text=black, inner sep=0pt]
  \tikzstyle{clientcommnum}=[commnum]

  \newcommand{\drawnodes}{
    \node[client] (c1) at (0, 2) {$c_1$};
    \node[client] (c2) at (0, 1) {$c_2$};
    \node[client] (c3) at (0, 0) {$c_3$};
    \node[proposer] (p1) at (1, 1.5) {$p_1$};
    \node[proposer] (p2) at (1, 0.5) {$p_2$};
    \node[matchmaker] (m1) at (2.00, 3.00) {$m_1$};
    \node[matchmaker] (m2) at (2.50, 2.75) {$m_2$};
    \node[matchmaker] (m3) at (3.00, 2.50) {$m_3$};
    \node[acceptor0] (a1) at (3, 1.75) {$a_1$};
    \node[acceptor0] (a2) at (3, 1.00) {$a_2$};
    \node[acceptor0] (a3) at (3, 0.25) {$a_3$};
    \node[acceptor1] (b1) at (2.00, -1.00) {$b_1$};
    \node[acceptor1] (b2) at (2.50, -0.75) {$b_2$};
    \node[acceptor1] (b3) at (3.00, -0.50) {$b_3$};
  }

  \begin{subfigure}[b]{0.3\textwidth}
    \centering
    \begin{tikzpicture}[xscale=1.5]
      \drawnodes{}
      \draw[comm, matchmakercolor, bend left=45] (p1) to node[commnum]{1} (m1);
      \draw[comm, matchmakercolor, bend left=25] (p1) to node[commnum]{1} (m2);
      \draw[comm, matchmakercolor, bend right=30] (m1) to node[commnum]{2} (p1);
      \draw[comm, matchmakercolor, bend right=10] (m2) to node[commnum]{2} (p1);
      \draw[clientcomm, bend left] (c1) to node[clientcommnum]{$a$} (p1);
      \draw[clientcomm, bend left=10] (p1) to node[clientcommnum]{$b$} (a2);
      \draw[clientcomm, bend right=10] (p1) to node[clientcommnum]{$b$} (a3);
      \draw[clientcomm, bend left=10] (a2) to node[clientcommnum]{$c$} (p1);
      \draw[clientcomm, bend left=25] (a3) to node[clientcommnum]{$c$} (p1);
      \draw[clientcomm, bend left] (p1) to node[clientcommnum]{$d$} (c1);
    \end{tikzpicture}
    \caption{Matchmaking}\figlabel{UnoptimizedReconfigurationMatchmaking}
  \end{subfigure}\hspace{0.04\textwidth}%
  \begin{subfigure}[b]{0.3\textwidth}
    \centering
    \begin{tikzpicture}[xscale=1.5]
      \drawnodes{}
      \draw[comm, config0color, bend left=20] (p1) to node[commnum]{3} (a1);
      \draw[comm, config0color, bend left=5] (p1) to node[commnum]{3} (a2);
      \draw[comm, config0color, bend left=0] (a1) to node[commnum]{4} (p1);
      \draw[comm, config0color, bend left=15] (a2) to node[commnum]{4} (p1);
      \draw[clientcomm] (c1) to node[clientcommnum]{$a$} (p1);
    \end{tikzpicture}
    \caption{Phase 1}\figlabel{UnoptimizedReconfigurationPhase1}
  \end{subfigure}\hspace{0.04\textwidth}%
  \begin{subfigure}[b]{0.3\textwidth}
    \centering
    \begin{tikzpicture}[xscale=1.5]
      \drawnodes{}
      \draw[comm, config1color, bend left=10] (p1) to node[commnum]{5} (b1);
      \draw[comm, config1color, bend left=20] (p1) to node[commnum]{5} (b2);
      \draw[comm, config1color, bend left=0] (b1) to node[commnum]{6} (p1);
      \draw[comm, config1color, bend right=8] (b2) to node[commnum]{6} (p1);
      \draw[clientcomm, bend left] (c1) to node[clientcommnum]{$a$} (p1);
      \draw[clientcomm, bend right=15] (p1) to node[clientcommnum]{$b$} (b1);
      \draw[clientcomm, bend left=50] (p1) to node[clientcommnum]{$b$} (b2);
      \draw[clientcomm, bend left=25] (b1) to node[clientcommnum]{$c$} (p1);
      \draw[clientcomm, bend right=35] (b2) to node[clientcommnum]{$c$} (p1);
      \draw[clientcomm, bend left] (p1) to node[clientcommnum]{$d$} (c1);
    \end{tikzpicture}
    \caption{Phase 2}\figlabel{UnoptimizedReconfigurationPhase2}
  \end{subfigure}
  \caption{%
    An example Matchmaker MultiPaxos reconfiguration without Optimization 2.
    The leader $p_1$ reconfigures from the acceptors $a_1$, $a_2$, $a_3$ to the
    acceptors $b_1$, $b_2$, $b_3$. Client commands are drawn as gray dashed
    lines. \markrevisions{Note that every subfigure shows one phase of a
    reconfiguration using solid colored lines, but the dashed lines show
    the complete execution of a client request that runs concurrently with the
    reconfiguration.} For simplicity, we assume that every proposer also serves
    as a replica.
  }%
  \figlabel{UnoptimizedReconfiguration}
\end{figure*}
}

This is illustrated in \figref{UnoptimizedReconfiguration}.
\figref{UnoptimizedReconfiguration} shows a leader $p_1$ reconfiguring from a
configuration of acceptors $C_\text{old}$ consisting of acceptors $a_1$, $a_2$,
and $a_3$ in round $i$ to a new configuration of acceptors $C_\text{new}$
consisting of acceptors $b_1$, $b_2$, and $b_3$ in round $i+1$. While the
leader performs the reconfiguration, clients continue to send state machine
commands to the leader. We consider such a command and perform a case analysis
on when the command arrives at the leader to see whether or not the command has
to be stalled.

\textbf{Case 1: Matchmaking (\figref{UnoptimizedReconfigurationMatchmaking})}.
If the leader receives a command during the Matchmaking phase, then the
leader can process the command as normal in round $i$ using the acceptors in
$C_\text{old}$. Even though the leader is executing the Matchmaking phase in
round $i+1$ and is communicating with the matchmakers, the acceptors in
$C_\text{old}$ are oblivious to this and can process commands in Phase $2$ in
round $i$.

\textbf{Case 2: Phase 1 (\figref{UnoptimizedReconfigurationPhase1})}.
If the leader receives a command during Phase 1, then the leader cannot process
the command. It must delay the processing of the command until Phase 1
finishes. Here's why. Once an acceptor in $C_\text{old}$ receives a
$\PhaseIA{i+1}$ message, it will reject any future commands in rounds less than
$i+1$, so the leader is unable to send the command to $C_\text{old}$. The
leader also cannot send the command to $C_\text{new}$ in round $i+1$ because it
has not yet finished executing Phase 1.
%

\textbf{Case 3: Phase 2 (\figref{UnoptimizedReconfigurationPhase2})}.
If the leader receives a command during Phase 2, then the leader can send the
command to the new acceptors in $C_\text{new}$ in round $i+1$. This is the
normal case of execution.

In summary, any commands received during Phase 1 of a reconfiguration are
delayed. Fortunately, we can eliminate this problem by applying Optimization 2
(Phase 1 Bypassing) from \secref{MatchmakerPaxos} to Matchmaker MultiPaxos.
Consider a leader performing a reconfiguration from $C_i$ in round $i$ to
$C_{i+1}$ in round $i+1$. At the end of the Matchmaking phase and at the
beginning of Phase 1 (in round $i+1$), let $k$ be the largest log entry that
the leader has assigned to a command. That is, all log entries after entry $k$
are empty. These log entries satisfy the preconditions of Optimization 2, so it
is safe for the leader to bypass Phase $1$ in round $i+1$ for these log entries
in the following way. When a leader receives a command after the Matchmaking
phase, it assigns the command a log entry larger than $k$, skips Phase 1, and
executes Phase $2$ in round $i+1$ with $C_\text{new}$ immediately.

With this optimization and the round scheme described in
\secref{Optimizations}, no state machine commands are delayed. Commands
received during the Matchmaking phase or earlier are chosen in round $i$ by
$C_\text{old}$ in log entries up to and including $k$. Commands received during
Phase 1, Phase 2, or later are chosen in round $i+1$ by $C_\text{new}$ in log
entries $k+1$, $k+2$, $k+3$, and so on. With this optimization Matchmaker
MultiPaxos can be reconfigured with minimal performance degradation.  We
confirm this empirically in \secref{Evaluation}.
}
{\section{Retiring Old Configurations}\seclabel{RetiringOldConfigurations}
We've discussed how to introduce new configurations. Now, we explain how to
shut down old configurations. We discuss Matchmaker Paxos, then Matchmaker
MultiPaxos.

\subsection{Matchmaker Paxos (How)}
At the beginning of round $i$, a proposer $p$ executes the Matchmaking phase and
computes a set $H_i$ of configurations in rounds less than $i$. The proposer
then executes Phase 1 with the acceptors in these configurations. Assume $H_i$
contains a configuration $C_j$ for a round $j < i$. If we prematurely shut down
the acceptors in $C_j$, then proposer $p$ will get stuck in Phase 1, waiting
for \textsc{Phase1B} messages from a quorum of nodes that have been shut down.
Therefore, we cannot shut down the acceptors in a configuration $C_j$ until we
are sure that the matchmakers will never again return $C_j$ during the
Matchmaking phase.

Thus, we extend Matchmaker Paxos to allow matchmakers to garbage collect
configurations from their logs, ensuring that the garbage collected
configurations will not be returned during any future Matchmaking phase.
More concretely, a proposer $p$ can now send a $\GarbageA{i}$ command to the
matchmakers informing them to garbage collect all configurations in rounds less
than $i$. When a matchmaker receives a $\GarbageA{i}$ message, it deletes log
entry $L[j]$ for every round $j < i$. It then updates a garbage collection
watermark $w$ to the maximum of $w$ and $i$ and sends back a $\GarbageB{i}$
message to the proposer. See \algoref{MatchmakerGcPseudocode}.

{\newcommand{\oldstuff}{\color{black!75}}
\newcommand{\newstuff}{\color{red}}

\begin{algorithm}
  \begin{algorithmic}[1]
    \oldstuff
    \GlobalState a log $L$ indexed by round, initially empty
    \newstuff
    \GlobalState \underline{a garbage collection watermark $w$, initially $0$}

    \Upon{\underline{receiving $\GarbageA{i}$ from proposer $p$}}
      \ForAll {\underline{$j < i$}}
        \State \underline{delete $L[j]$}
      \EndFor
      \State \underline{$w \gets \max(w, i)$}
      \State \underline{send $\GarbageB{i}$ to $p$}
    \EndUpon

    \oldstuff
    \Upon{receiving $\MatchA{i}{C_i}$ from proposer $p$}
      \If{{\newstuff \underline{$i < w$}} or $\exists$ $C_j$ in round $j \geq i$ in $L$}
        \State ignore the $\MatchA{i}{C_i}$ message
      \Else
        \State $H_i \gets \setst{(j, C_j)}{C_j \in L}$
        \State $L[i] \gets C_i$
        \State send $\MatchBGc{i}{{\newstuff \underline{w}}}{H_i}$ to $p$
      \EndIf
    \EndUpon
  \end{algorithmic}
  \caption{%
    Matchmaker Pseudocode (with GC). Changes to \algoref{MatchmakerPseudocode}
    are shown in red.
  }%
  \algolabel{MatchmakerGcPseudocode}
\end{algorithm}

}

We also update the Matchmaking phase in three ways.
First, a matchmaker ignores a $\MatchA{i}{C_i}$ message if $i$ has been garbage
collected (i.e.\ if $i < w$).
Second, a matchmaker returns its garbage collection watermark $w$ in every
\textsc{MatchB} that it sends.
Third, when a proposer receives $\MatchBGc{i}{w_1}{H_i^1}$, $\ldots$,
$\MatchBGc{i}{w_{f+1}}{H_i^{f+1}}$ from $f+1$ matchmakers, it again computes
$H_i = \cup_{j=1}^{f+1} H_i^{j}$. It then computes $w = \max_{j=1}^{f+1} w_j$
and prunes every configuration in $H_i$ in a round less than $w$. In other
words, if any of the $f+1$ matchmakers have garbage collected round $j$, then
the proposer also garbage collects round $j$.

Once a proposer receives $\GarbageB{i}$ messages from at least $f+1$
matchmakers $M$, it is guaranteed that all future Matchmaking phases will not
include any configuration in any round less than $i$. Why? Consider a future
Matchmaking phase run with $f+1$ matchmakers $M'$. $M$ and $M'$ intersect, so
some matchmaker $m$ in the intersection has a garbage collection watermark at
least as large as $i$.

Thus, once a configuration has been garbage collected by a majority of the
matchmakers, we can shut down the acceptors in the configuration.

\subsection{Matchmaker Paxos (When)}
Once a configuration has been garbage collected, it is safe to shut it down,
but when is it safe to garbage collect a configuration? It is certainly not
always safe. For example, if we prematurely garbage collect configuration $C_j$
in round $j$, a future proposer in round $i > j$ may not learn about a value $v$
chosen in round $j$ and then erroneously get a value other than $v$ chosen in
round $i$. There are three situations in which it is safe for a proposer $p_i$
in round $i$ to issue a $\GarbageA{i}$ command. We explain the three situations
and provide intuition on why they are safe.
\iftoggle{techreportenabled}{}{%
  We refer to the technical report for a proof~\cite{whittaker2020matchmaker}.
}

\textbf{Scenario 1.}
If the proposer $p_i$ gets a value $x$ chosen in round $i$, then it can safely
issue a $\GarbageA{i}$ command. Why?
When a proposer $p_j$ in round $j > i$ executes Phase 1, it will learn
about the value $x$ and propose $x$ in Phase 2. But first, it must establish
that no value other than $x$ has been or will be chosen in any round less than
$j$. The proposer $p_i$ already established this fact for all rounds less than
$i$, so any communication with the configurations in these rounds is redundant.
Thus, we can garbage collect them.



\textbf{Scenario 2.}
If the proposer $p_i$ executes Phase $1$ in round $i$ and finds $k = -1$, then
it can safely issue a $\GarbageA{i}$ command. Recall from
\secref{MatchmakerPaxos} that if $k=-1$, then no value has been or will be
chosen in any round less than $i$. This situation is similar to Scenario 1. Any
future proposer $p_j$ in round $j > i$ does not have to redundantly communicate
with the configurations in rounds less than $i$ since $p_i$ already established
that no value has been chosen in these rounds.

\textbf{Scenario 3.}
If the proposer $p_i$ learns that a value $x$ has already been chosen and has
been stored on $f+1$ other machines, then the proposer can safely issue a
$\GarbageA{i}$ command after it informs a Phase 2 quorum of acceptors in $C_i$
of this fact. Any future proposer $p_j$ in round $j > i$ will contact a Phase 1
quorum of $C_i$ and encounter at least one acceptor that knows the value $x$
has already been chosen. When this acceptor informs $p_j$ that a value $x$ has
already been chosen, $p_j$ stops executing the protocol entirely and simply
fetches the value $x$ from one of the $f+1$ machines that store the value.
\begin{revisions}
  Note that the decision of exactly which $f+1$ machines is not important. Any
  $f+1$ machines can be used. In our implementation, we use $f+1$ replicas.
\end{revisions}

\begin{techreport}

\newcommand{\newstuff}[1]{\textcolor{red}{#1}}

To prove that these three scenarios are safe, we repeat the safety proof from
\secref{MatchmakerPaxos}. The new bits are shown in red.

\begin{proof}
We prove, for every round $i$, the statement $P(i)$: ``if a proposer proposes a
value $v$ in round $i$ (i.e.\ sends a \textsc{Phase2A} message for value $v$ in
round $i$), then no value other than $v$ has been or will be chosen in any
round less than $i$.'' $P(i)$ suffices to prove that Matchmaker Paxos is safe
for the following reason. Assume for contradiction that Matchmaker Paxos
chooses distinct values $x$ and $y$ in rounds $i$ and $j$ with $i < j$.  Some
proposer must have proposed $y$ in round $j$, so $P(j)$ ensures us that no
value other than $y$ could have been chosen in round $i$. But, $x$ was chosen,
a contradiction.

We prove $P(i)$ by strong induction on $i$. $P(0)$ is vacuous because there are
no rounds less than $0$. For the general case $P(i)$, we assume $P(0), \ldots,
P(i-1)$. We perform a case analysis on the proposer's pseudocode. Either $k$ is
$-1$ or it is not (line 11). First, assume it is not. In this case, the
proposer proposes $x$, the value proposed in round $k$ (line 12). We perform a
case analysis on $j$ to show that no value other than $x$ has been or will be
chosen in any round $j < i$.

\textbf{Case 1: $j > k$.}
We show that no value has been or will be chosen in round $j$.
Recall that at the end of the Matchmaking phase, the proposer computed the
set $H_i$ of prior configurations using responses from a set $M$ of $f+1$
matchmakers. Either $H_i$ contains a configuration $C_j$ in round $j$ or it
doesn't.

First, suppose it does. Then, the proposer sent $\PhaseIA{i}$ messages to all
of the acceptors in $C_j$. A Phase 1 quorum of these acceptors, say $Q$, all
received $\PhaseIA{i}$ messages and replied with \textsc{Phase1B} messages.
Thus, every acceptor in $Q$ set its round $r$ to $i$, and in doing so, promised
to never vote in any round less than $i$.  Moreover, none of the acceptors in
$Q$ had voted in any round greater than $k$. So, every acceptor in $Q$ has not
voted and never will vote in round $j$. For a value $v'$ to be chosen in round
$j$, it must receive votes from some Phase 2 quorum $Q'$ of round $j$
acceptors. But, $Q$ and $Q'$ necessarily intersect, so this is impossible.
Thus, no value has been or will be chosen in round $j$.

Now suppose that $H_i$ does \emph{not} contain a configuration for round $j$.
\newstuff{Either a configuration $C_j$ was garbage collected from $H_i$ or it
wasn't. First, assume it wasn't.} Then, $H_i$ is the union of $f+1$
$\textsc{MatchB}$ messages from the $f+1$ matchmakers in $M$. Thus, if $H_i$
does not contain a configuration for round $j$, then none of the
$\textsc{MatchB}$ messages did either. This means that for every matchmaker $m
\in M$, when $m$ received $\MatchA{i}{C_i}$, it did not contain a configuration
for round $j$ in its log \newstuff{and never did}. Moreover, by processing the
$\MatchA{i}{C_i}$ request and inserting $C_i$ in log entry $i$, the matchmaker
is guaranteed to never process a $\MatchA{j}{C_j}$ request in the future. Thus,
every matchmaker in $M$ has not processed a \textsc{MatchA} request in round
$j$ and never will. For a value to be chosen in round $j$, the proposer
executing round $j$ must first receive replies from $f+1$ matchmakers, say
$M'$, in round $j$. But, $M$ and $M'$ necessarily intersect, so this is
impossible. Thus, no value has been or will be chosen in round $j$.

\newstuff{%
Otherwise, a configuration $C_j$ was garbage collected from $H_i$. Note that
none of the matchmakers in $M$ had received a $\GarbageA{i'}$ command for a
round $i' > i$ when they responded with their \textsc{MatchB} messages. If they
had, they would have ignored our $\MatchA{i}{C_i}$ message. Let $i'$ be the
largest round $j < i' < i$ such that a matchmaker in $M$ had received a
$\GarbageA{i'}$ message before responding to our $\MatchA{i}{C_i}$ message.
}

\newstuff{%
If $i'$ was garbage collected because of Scenario 1, then $k$ would be at least
as large as $i'$ since we would have intersected the Phase $2$ quorum of
$C_{i'}$ used in round $i'$ to get a value chosen. But $k < j < i'$, a
contradiction. If $i'$ was garbage collected because of Scenario 2, then we
know no value has been or will be chosen in round $j$. If $i'$ was garbage
collected because of Scenario 3, then we would have intersected the Phase $2$
quorum of $C_{i'}$ that knows a value was already chosen, and we would have not
proposed a value in the first place. But, we proposed $x$, a contradiction.
}

\textbf{Case 2: $j = k$.}
In a given round, at most one value is proposed, let alone chosen. $x$ is
\emph{the} value proposed in round $k$, so no other value could be chosen in
round $k$.

\textbf{Case 3: $j < k$.}
  We can apply the inductive hypothesis to get $P(k)$ which states that no
  value other than $x$ has been or will be chosen in any round less than $k$.
  This includes round $j$, which is exactly what we're trying to prove.

Finally, if $k$ is $-1$, then we are in the same situation as in Case 1.
\end{proof}
\end{techreport}

\subsection{Matchmaker MultiPaxos}
Recall that the Matchmaker MultiPaxos leader $p_i$ in round $i$ uses a single
configuration $C_i$ for \emph{every} log entry. The leader $p_i$ can safely
issue a $\GarbageA{i}$ command to the matchmakers once it ensures that
\emph{every} log entry satisfies one of the three scenarios listed above. It
does so as follows.

Recall from \figref{MultiPaxosPhase1Log} that at the end of Phase 1 and
at the beginning of Phase 2, the log can be divided into three regions: a
prefix of log entries that the leader knows have already been chosen, a
subsequence of commands that the leader gets chosen in Phase 2, and an infinite
tail of empty entries.
Scenario 2 applies to the infinite tail of empty log entries. These are the log
entries for which $k=-1$. Scenario 1 applies to the middle subsequence of
commands once the leader gets them chosen. Scenario 3 applies to the prefix of
chosen entries if we make the following adjustments. First, we deploy $2f+1$
replicas instead of $f+1$. Second, the leader ensures that the prefix of
previously chosen log entries are stored on at least $f+1$ of the $2f+1$
replicas (this ensures that despite $f$ replica failures, some replica will
store the values). Third, the leader informs a Phase 2 quorum of $C_i$
acceptors that these commands have been stored on the replicas.

In summary, the leader $p_i$ of round $i$ executes as follows. It executes the
Matchmaking phase yielding prior configurations $H_i$. It then executes Phase 1
with the configurations in $H_i$. It enters Phase 2 and gets the middle
subsequence of commands chosen. It also informs a Phase 2 quorum of $C_i$
acceptors once the prefix of previously chosen commands have been stored on
$f+1$ replicas. It then issues a $\GarbageA{i}$ command to the matchmakers and
awaits $f+1$ $\GarbageB{i}$ responses. At this point, all previous
configurations can be shut down.

Note that the leader can begin processing state machine commands from clients
as soon as it enters Phase 2. It does not have to stall client commands during
garbage collection. Note also that during normal operation (i.e.\ operation
with a single stable leader), old configurations are garbage collected after a
few round trips of communication. In \secref{Evaluation}, we find that $H_i$
almost always contains only a single configuration (i.e.\ $C_{i-1}$).
}
{\section{Reconfiguring Matchmakers}\seclabel{ReconfiguringMatchmakers}
We've discussed how Matchmaker MultiPaxos allows us to reconfigure the set of
acceptors. In this section, we discuss how to reconfigure proposers, replicas,
and matchmakers.

Reconfiguring proposers and replicas is straightforward.
In fact, Matchmaker MultiPaxos reconfigures proposers and replicas in exactly
the same way as MultiPaxos~\cite{van2015paxos}, so we do not discuss it at
length. In short, a proposer can be safely added or removed at any time.
Replicas can also be safely added or removed at any time so long as we ensure
that commands replicated on $f+1$ replicas remain replicated on $f+1$ replicas.
For performance, a newly introduced proposer should contact an existing
proposer or replica to learn about the prefix of already chosen commands, and a
newly introduced replica should copy the log from an existing replica.

Reconfiguring matchmakers is a bit more involved, but still relatively
straightforward. Recall that proposers performs the Matchmaking phase only
during a \markrevisions{change in round}. Thus, for the vast majority of the
time---specifically, when there is a single, stable leader---the matchmakers
are idle. When idle, they don't send or receive any messages and are off the
critical path of command processing. This means that the way we reconfigure the
matchmakers has to be safe, but it doesn't have to be efficient. The
matchmakers can be reconfigured at any time between leader changes without any
impact on the performance of the protocol.

Thus, we use the simplest approach to reconfiguration: we shut down the old
matchmakers and replace them with new ones, making sure that the new
matchmakers' initial state is the same as the old matchmakers' final state.
More concretely, we reconfigure from a set $M_\text{old}$ of matchmakers to a
new set $M_\text{new}$ as follows. First, a proposer (or any other node, it
doesn't really matter) sends a $\StopA{}$ message to the matchmakers in
$M_\text{old}$. When a matchmaker $m_i$ receives a $\StopA{}$ message, it stops
processing messages (except for other $\StopA{}$ messages) and replies with
$\StopB{L_i}{w_i}$ where $L_i$ is $m_i$'s log and $w_i$ is its garbage
collection watermark. When the proposer receives $\textsc{StopB}$ messages from
$f+1$ matchmakers, it knows that the matchmakers have effectively been
shut down. It computes $w$ as the maximum of every returned $w_i$. It computes
$L$ as the union of the returned logs, and removes all entries of $L$ that
appear in a round less than $w$.
\begin{techreport}
  An example of this log merging is illustrated in
  \figref{MergingMatchmakerLogs}.
\end{techreport}

\iftoggle{techreportenabled}{\begin{figure}[ht]
  \centering
  \tikzstyle{logentry}=[draw, line width=1pt, minimum width=17pt,
                        minimum height=17pt]
  \tikzstyle{lognum}=[gray, font=\scriptsize]
  \tikzstyle{garbage}=[fill=flatred!75]
  \begin{tikzpicture}[xscale=1.5]
    \node[logentry, label={[lognum]180:0}] (a0) at (0, 0) {$C_0$};
    \node[logentry, label={[lognum]180:1}, above=-1pt of a0] (a1) {$C_1$};
    \node[logentry, label={[lognum]180:2}, above=-1pt of a1] (a2) {};
    \node[logentry, label={[lognum]180:3}, above=-1pt of a2] (a3) {};
    \node[logentry, label={[lognum]180:4}, above=-1pt of a3] (a4) {$C_4$};
    \node[below=0pt of a0] {$L_0$};

    \node[logentry, garbage, label={[lognum]180:0}] (b0) at (1, 0) {$\times$};
    \node[logentry, garbage, label={[lognum]180:1}, above=-1pt of b0] (b1) {$\times$};
    \node[logentry, label={[lognum]180:2}, above=-1pt of b1] (b2) {$C_2$};
    \node[logentry, label={[lognum]180:3}, above=-1pt of b2] (b3) {};
    \node[logentry, label={[lognum]180:4}, above=-1pt of b3] (b4) {};
    \node[below=0pt of b0] {$L_1$};

    \node[logentry, garbage, label={[lognum]180:0}] (c0) at (2, 0) {$\times$};
    \node[logentry, label={[lognum]180:1}, above=-1pt of c0] (c1) {$C_1$};
    \node[logentry, label={[lognum]180:2}, above=-1pt of c1] (c2) {$C_2$};
    \node[logentry, label={[lognum]180:3}, above=-1pt of c2] (c3) {};
    \node[logentry, label={[lognum]180:4}, above=-1pt of c3] (c4) {};
    \node[below=0pt of c0] {$L_2$};

    \draw[-latex, line width=2pt] (c2.east) ++ (0.25, 0) to ++(1, 0);

    \node[logentry, garbage, label={[lognum]180:0}] (d0) at (4, 0) {$\times$};
    \node[logentry, garbage, label={[lognum]180:1}, above=-1pt of d0] (d1) {$\times$};
    \node[logentry, label={[lognum]180:2}, above=-1pt of d1] (d2) {$C_2$};
    \node[logentry, label={[lognum]180:3}, above=-1pt of d2] (d3) {};
    \node[logentry, label={[lognum]180:4}, above=-1pt of d3] (d4) {$C_4$};
  \end{tikzpicture}
  \caption{%
    An example of merging three matchmaker logs ($L_0$, $L_1$, and $L_2$)
    during a matchmaker reconfiguration.  Garbage collected log entries are
    shown in red.
  }
  \figlabel{MergingMatchmakerLogs}
\end{figure}
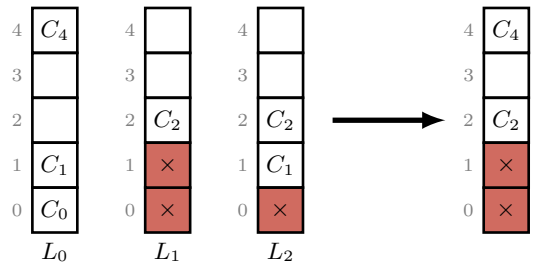}{}

The proposer then sends $L$ and $w$ to all of the matchmakers in
$M_\text{new}$. Each matchmaker adopts these values as its initial state. At
this point, the matchmakers in $M_\text{new}$ \emph{cannot} begin processing
commands yet. Naively, it is possible that two different nodes could
simultaneously attempt to reconfigure to two disjoint sets of matchmakers, say
$M_\text{new}$ and $M_\text{new}'$. To avoid this, every matchmaker in
$M_\text{old}$ doubles as a Paxos acceptor. A proposer attempting to
reconfigure to $M_\text{new}$ acts as a Paxos proposer and gets the value
$M_\text{new}$ chosen by the matchmakers (which are acting as Paxos acceptors).
Once $M_\text{new}$ is chosen, the proposer notifies the matchmakers in
$M_\text{new}$ that the reconfiguration is complete and that they are free to
start processing commands. A safety proof for Matchmaker MultiPaxos with
matchmaker reconfiguration is given in the technical
report~\cite{whittaker2020matchmaker}.

\begin{techreport}

\newcommand{\newstuff}[1]{\textcolor{red}{#1}}

We repeat the safety proof from \secref{RetiringOldConfigurations}. The new
bits are shown in red.

\begin{proof}
We prove, for every round $i$, the statement $P(i)$: ``if a proposer proposes a
value $v$ in round $i$ (i.e.\ sends a \textsc{Phase2A} message for value $v$ in
round $i$), then no value other than $v$ has been or will be chosen in any
round less than $i$.'' $P(i)$ suffices to prove that Matchmaker Paxos is safe
for the following reason. Assume for contradiction that Matchmaker Paxos
chooses distinct values $x$ and $y$ in rounds $i$ and $j$ with $i < j$.  Some
proposer must have proposed $y$ in round $j$, so $P(j)$ ensures us that no
value other than $y$ could have been chosen in round $i$. But, $x$ was chosen,
a contradiction.

We prove $P(i)$ by strong induction on $i$. $P(0)$ is vacuous because there are
no rounds less than $0$. For the general case $P(i)$, we assume $P(0), \ldots,
P(i-1)$. We perform a case analysis on the proposer's pseudocode. Either $k$ is
$-1$ or it is not (line 11). First, assume it is not. In this case, the
proposer proposes $x$, the value proposed in round $k$ (line 12). We perform a
case analysis on $j$ to show that no value other than $x$ has been or will be
chosen in any round $j < i$.

\textbf{Case 1: $j > k$.}
We show that no value has been or will be chosen in round $j$.
Recall that at the end of the Matchmaking phase, the proposer computed the
set $H_i$ of prior configurations using responses from a set $M$ of $f+1$
matchmakers. Either $H_i$ contains a configuration $C_j$ in round $j$ or it
doesn't.

First, suppose it does. Then, the proposer sent $\PhaseIA{i}$ messages to all
of the acceptors in $C_j$. A Phase 1 quorum of these acceptors, say $Q$, all
received $\PhaseIA{i}$ messages and replied with \textsc{Phase1B} messages.
Thus, every acceptor in $Q$ set its round $r$ to $i$, and in doing so, promised
to never vote in any round less than $i$.  Moreover, none of the acceptors in
$Q$ had voted in any round greater than $k$. So, every acceptor in $Q$ has not
voted and never will vote in round $j$. For a value $v'$ to be chosen in round
$j$, it must receive votes from some Phase 2 quorum $Q'$ of round $j$
acceptors. But, $Q$ and $Q'$ necessarily intersect, so this is impossible.
Thus, no value has been or will be chosen in round $j$.

Now suppose that $H_i$ does \emph{not} contain a configuration for round $j$.
Either a configuration $C_j$ was garbage collected from $H_i$ or it
wasn't. First, assume it wasn't. Then, $H_i$ is the union of $f+1$
$\textsc{MatchB}$ messages from the $f+1$ matchmakers in $M$. Thus, if $H_i$
does not contain a configuration for round $j$, then none of the
$\textsc{MatchB}$ messages did either. This means that for every matchmaker $m
\in M$, when $m$ received $\MatchA{i}{C_i}$, it did not contain a configuration
for round $j$ in its log and never did. \newstuff{Moreover, no majority in any
previous set of matchmakers contained a configuration in round $j$. If any
majority did have a configuration in round $j$, then all subsequent matchmakers
would as well since a set of matchmakers is initialized from a majority of the
previous matchmakers.} Moreover, by processing the $\MatchA{i}{C_i}$ request
and inserting $C_i$ in log entry $i$, the matchmaker is guaranteed to never
process a $\MatchA{j}{C_j}$ request in the future. \newstuff{Moreover, no
future set of matchmakers will either. A majority of $M$ have a configuration in
entry $i$, so all subsequent configurations will as well. Therefore, they will
all reject a configuration in round $j$.} Thus, every matchmaker in $M$ has not
processed a \textsc{MatchA} request in round $j$ and never will. For a value to
be chosen in round $j$, the proposer executing round $j$ must first receive
replies from $f+1$ matchmakers, say $M'$, in round $j$. But, $M$ and $M'$
necessarily intersect, so this is impossible. \newstuff{This argument holds for
every set of matchmakers.} Thus, no value has been or will be chosen in round
$j$.

Otherwise, a configuration $C_j$ was garbage collected from $H_i$. Note that
none of the matchmakers in $M$ had received a $\GarbageA{i'}$ command for a
round $i' > i$ when they responded with their \textsc{MatchB} messages. If they
had, they would have ignored our $\MatchA{i}{C_i}$ message.
\newstuff{Similarly, none of the matchmakers in $M$ were initialized with a
garbage collection watermark $w > i$.} Let $i'$ be the largest round $j < i' <
i$ that a matchmaker in $M$ \newstuff{garbage collected} before responding to
our $\MatchA{i}{C_i}$ message.

If $i'$ was garbage collected because of Scenario 1, then $k$ would be at least
as large as $i'$ since we would have intersected the Phase $2$ quorum of
$C_{i'}$ used in round $i'$ to get a value chosen. But $k < j < i'$, a
contradiction. If $i'$ was garbage collected because of Scenario 2, then we
know no value has been or will be chosen in round $j$. If $i'$ was garbage
collected because of Scenario 3, then we would have intersected the Phase $2$
quorum of $C_{i'}$ that knows a value was already chosen, and we would have not
proposed a value in the first place. But, we proposed $x$, a contradiction.

\textbf{Case 2: $j = k$.}
In a given round, at most one value is proposed, let alone chosen. $x$ is
\emph{the} value proposed in round $k$, so no other value could be chosen in
round $k$.

\textbf{Case 3: $j < k$.}
We can apply the inductive hypothesis to get $P(k)$ which states that no
value other than $x$ has been or will be chosen in any round less than $k$.
This includes round $j$, which is exactly what we're trying to prove.

Finally, if $k$ is $-1$, then we are in the same situation as in Case 1.
\end{proof}
\end{techreport}
}
{\section{Insights and Generality}\seclabel{Generality}

\subsection{Insights}
\textbf{Vertical Paxos}
Matchmaker Paxos was inspired by Vertical Paxos~\cite{lamport2009vertical}, but
differs in ways that improve simplicity, efficiency, and garbage collection.
In terms of simplicity, Vertical Paxos requires an external master, which is
itself implemented using state machine replication. Our matchmakers are
analogous to the external master but show that such a master does not require a
nested invocation of state machine replication.
From an efficiency perspective, Vertical Paxos requires that a proposer
\markrevisions{executes} Phase 1 in order to perform a reconfiguration. Thus,
Vertical Paxos \markrevisions{cannot} be extended to MultiPaxos without causing
performance degradation during reconfiguration. This is not the case for
matchmakers thanks to Phase 1 Bypassing.
Finally, Vertical Paxos garbage collects old configurations in situations
similar to Scenario 1 and Scenario 2 from \secref{RetiringOldConfigurations}.
It does not include Scenario 3, and as a result, it is unclear how to extend
Vertical Paxos' garbage collection to MultiPaxos.


%

\textbf{Fast Paxos.}
Fast Paxos~\cite{lamport2006fast} is a Paxos variant that shaves off one
network delay from Paxos in the best case, but can have higher delays if
concurrently proposed commands conflict. While Paxos quorums consist of $f+1$
out of $2f+1$ acceptors, Fast Paxos requires larger quorums. Many protocols
have reduced Fast Paxos quorum sizes a bit, but to date, Fast Paxos quorum
sizes have remained larger than classic Paxos quorum sizes.
Using matchmakers, we can implement Fast Paxos with a fixed set of $f+1$
acceptors (and hence with $f+1$-sized quorums).
%
%
Specifically, we deploy Fast Paxos with $f+1$ acceptors, with a single
unanimous Phase 2 quorum, and with singleton Phase 1 quorums. A full
description of the protocol and a proof of correctness is given in our
technical report~\cite{whittaker2020matchmaker}.



\iftoggle{techreportenabled}{%
  \begin{algorithm}[t]
  \begin{algorithmic}[1]
    \GlobalState a round $i$, initially $-1$
    \GlobalState the configuration $C_i$ for round $i$, initially \textsf{null}
    \GlobalState the prior configurations $H_i$ for round $i$, initially \textsf{null}

    \State $i \gets$ next largest round owned by this proposer
    \State $C_i \gets$ an arbitrary configuration
    \State send $\MatchA{i}{C_i}$ to all of the matchmakers

    \Upon{%
      receiving $\MatchB{i}{H_i^1}, \ldots, \MatchB{i}{H_i^{f+1}}$ from $f+1$
      matchmakers
    }
      \State $H_i \gets \bigcup_{j=1}^{f+1} H_i^j$
      \State send $\PhaseIA{i}$ to every acceptor in $H_i$
    \EndUpon

    \Upon{%
      receiving $\PhaseIB{i}{-}{-}$ from a Phase 1 quorum from every
      configuration in $H_i$
    }
      \State $k \gets$ the largest $vr$ in any $\PhaseIB{i}{vr}{vv}$
      \State $V \gets$ the corresponding $vv$'s in round $k$
      \If{$k = -1$}
        \State send $\PhaseIIA{i}{\textsf{any}}$ to every acceptor in $C_i$
      \ElsIf{$V = \set{v}$}
        \State send $\PhaseIIA{i}{v}$ to every acceptor in $C_i$
      \Else{}
        \State send $\PhaseIIA{i}{\textsf{any}}$ to every acceptor in $C_i$
      \EndIf
    \EndUpon
  \end{algorithmic}
  \caption{Fast Paxos Proposer Pseudocode}%
  \algolabel{FastPaxosProposerPseudocode}
\end{algorithm}

}{}

\begin{techreport}

Fast Paxos proposer pseudocode is given in
\algoref{FastPaxosProposerPseudocode}. We do not modify the Fast Paxos acceptor
or the matchmakers. For simplicity, we assume that we are using the unanimous
configurations described above. Generalizing to arbitrary configurations that
satisfy Fast Paxos' quorum intersection requirements is straightforward. Note
that Fast Paxos cannot leverage Phase 1 Bypassing. Also note while both
MultiPaxos and our Fast Paxos variant both have quorums of size $f+1$, our Fast
Paxos variant has a \emph{fixed} set of $f+1$ acceptors, while MultiPaxos can
choose any set of $f + 1$ acceptors from all $2f+1$ acceptors. This has some
disadvantages in terms of tail latency and fault tolerance.

We now prove that our modifications to Fast Paxos are safe. For simplicity, we
ignore garbage collection and matchmaker reconfiguration. Introducing those two
features and proving them correct is pretty much identical to what we did with
Matchmaker Paxos.

\begin{proof}
We prove, for every round $i$, the statement $P(i)$ which states that if a an
acceptor votes for a value $v$ in round $i$ (i.e.\ sends a \textsc{Phase2B}
message for value $v$ in round $i$), then no value other than $v$ has been or
will be chosen in any round less than $i$. $P(i)$ suffices to prove that
Matchmaker Paxos is safe. Why? Well, assume for contradiction that Matchmaker
Paxos chooses distinct values $x$ and $y$ in rounds $i$ and $j$ with $i < j$.
Some acceptor must have voted for $y$ in round $j$, so $P(j)$ ensures us that
no value other than $y$ could have been chosen in round $i$. But, $x$ was
chosen, a contradiction.

We prove $P(i)$ by strong induction on $i$. $P(0)$ is vacuous because there are
no rounds less than $0$. For the general case $P(i)$, we assume $P(0), \ldots,
P(i-1)$. We perform a case analysis on the proposer's pseudocode. Either $k$ is
$-1$ or it is not (line 8). First, assume it is not. We perform a case
analysis on rounds $j < i$.

\textbf{Case 1: $j > k$.}
Recall that at the end of the Matchmaking phase, the proposer computed the
set $H_i$ of prior configurations using responses from a set $M$ of $f+1$
matchmakers. Either $H_i$ contains a configuration $C_j$ in round $j$ or it
doesn't.

First, suppose it does. Then, the proposer sent $\PhaseIA{i}$ messages to all
of the acceptors in $C_j$. A Phase 1 quorum of these acceptors, say $Q$, all
received $\PhaseIA{i}$ messages and replied with \textsc{Phase1B} messages.
Thus, every acceptor in $Q$ set its round $r$ to $i$, and in doing so, promised
to never vote in any round less than $i$.  Moreover, none of the acceptors in
$Q$ had voted in any round greater than $k$. So, every acceptor in $Q$ has not
voted and never will vote in round $j$. For a value $v'$ to be chosen in round
$j$, it must receive votes from some Phase 2 quorum $Q'$ of round $j$
acceptors. But, $Q$ and $Q'$ necessarily intersect, so this is impossible.
Thus, no value has been or will be chosen in round $j$.

Now suppose that $H_i$ does \emph{not} contain a configuration for round $j$.
$H_i$ is the union of $f+1$ $\textsc{MatchB}$ messages from the $f+1$
matchmakers in $M$. Thus, if $H_i$ does not contain a configuration for round
$j$, then none of the $\textsc{MatchB}$ messages did either. This means that
for every matchmaker $m \in M$, when $m$ received $\MatchA{i}{C_i}$, it did not
contain a configuration for round $j$ in its log. Moreover, by processing the
$\MatchA{i}{C_i}$ request and inserting $C_i$ in log entry $i$, the matchmaker
is guaranteed to never process a $\MatchA{j}{C_j}$ request in the future. Thus,
every matchmaker in $M$ has not processed a \textsc{MatchA} request in round
$j$ and never will. For a value to be chosen in round $j$, the proposer
executing round $j$ must first receive replies from $f+1$ matchmakers, say
$M'$, in round $j$. But, $M$ and $M'$ necessarily intersect, so this is
impossible. Thus, no value has been or will be chosen in round $j$.

\textbf{Case 2: $j = k$.}
If $V = \set{v}$, then the proposer proposes $v$. We must prove that no value
other than $v$ has been or will be chosen in round $k$. For a value to be
chosen in round $k$, every acceptor must vote for it in round $k$. Some
acceptor voted for $v$ in round $k$, so it is the only value with the
possibility of receiving a unanimous vote.

Otherwise $V$ contains multiple distinct elements, and the proposer proposes
\textsf{any}. We must prove that no value has been or will be chosen in round
$k$. This is immediate since no value can receive a unanimous vote in round
$k$, if two different values have received votes in round $k$.

\textbf{Case 3: $j < k$.}
If $V = \set{v}$, then the proposer proposes $v$, and we must prove that no
value other than $v$ has been or will be chosen in any round less than $k$.
This is immediate from $P(k)$. Otherwise, $V = \set{v_1, v_2, \ldots}$, and the
proposer proposes \textsf{any}. We must prove that no value has been or will be
chosen in any round less than $k$. $P(k)$ tells us that no value other than
$v_1$ has been or will be chosen in any round less than $k$.  $P(k)$ also tells
us that no value other than $v_2$ has been or will be chosen in any round less
than $k$.  Thus, no value has been or will be chosen in any round less than
$k$.

Finally, if $k$ is $-1$, then we are in the same situation as in Case 1. No
value has been or will be chosen in any round less than $i$.
\end{proof}
\end{techreport}

\textbf{DPaxos.}
Matchmaker Paxos subsumes DPaxos and corrects some previously undiscovered
errors in the protocol. DPaxos is a Paxos variant that allows every round to
use a different subset of acceptors from some fixed set of acceptors.
Matchmaker Paxos obviates the need for a fixed set of nodes. DPaxos' scope is
limited to a single instance of consensus, whereas Matchmaker MultiPaxos shows
how to efficiently reconfigure across multiple instances of consensus
simultaneously. We also discovered that DPaxos' garbage collection algorithm is
unsafe. Matchmaker MultiPaxos' garbage collection algorithm fixes the bug (see
our technical report for details~\cite{whittaker2020matchmaker}).

\begin{techreport}

\newcommand{\readQuorum}{leader election quorum}
\newcommand{\ReadQuorum}{Leader election quorum}
\newcommand{\readPhase}{leader election phase}
\newcommand{\ReadPhase}{Leader election phase}

\newcommand{\writeQuorum}{replication quorum}
\newcommand{\WriteQuorum}{Replication quorum}
\newcommand{\writePhase}{replication phase}
\newcommand{\WritePhase}{replication phase}

Consider a DPaxos deployment with $f_d = 1$, $f_z = 0$, three zones, three
nodes per zone, and delegate quorums. Thus, a \writeQuorum{} consists of two
nodes in one zone, and a \readQuorum{} consists of two nodes in two zones. We
name the nodes $A$ through $I$. Beside each node, we display its ballot, vote
ballot, vote value, and intents~\cite{lamport2001paxos}.

\tikzstyle{proc}=[draw, circle, inner sep=0pt]
\tikzstyle{zone1}=[fill=flatred!25]
\tikzstyle{zone2}=[fill=flatgreen!25]
\tikzstyle{zone3}=[fill=flatblue!25]
\newcommand{\nodes}{
    \node (zone1) at (0, 3) {Zone 1};
    \node[proc, zone1] (A) at (0, 2) {$A$};
    \node[proc, zone1] (B) at (0, 1) {$B$};
    \node[proc, zone1] (C) at (0, 0) {$C$};
    \node (zone2) at (2, 3) {Zone 2};
    \node[proc, zone2] (D) at (2, 1.5) {$D$};
    \node[proc, zone2] (E) at (2, 0.5) {$E$};
    \node[proc, zone2] (F) at (2, -0.5) {$F$};
    \node (zone3) at (4, 3) {Zone 3};
    \node[proc, zone3] (G) at (4, 2) {$G$};
    \node[proc, zone3] (H) at (4, 1) {$H$};
    \node[proc, zone3] (I) at (4, 0) {$I$};
}

\begin{center}
  \begin{tikzpicture}[xscale=1]
    \nodes
    \node[right=0cm of A] {\small $-1, -1, \bot, \emptyset$};
    \node[right=0cm of B] {\small $-1, -1, \bot, \emptyset$};
    \node[right=0cm of C] {\small $-1, -1, \bot, \emptyset$};
    \node[right=0cm of D] {\small $-1, -1, \bot, \emptyset$};
    \node[right=0cm of E] {\small $-1, -1, \bot, \emptyset$};
    \node[right=0cm of F] {\small $-1, -1, \bot, \emptyset$};
    \node[right=0cm of G] {\small $-1, -1, \bot, \emptyset$};
    \node[right=0cm of H] {\small $-1, -1, \bot, \emptyset$};
    \node[right=0cm of I] {\small $-1, -1, \bot, \emptyset$};
  \end{tikzpicture}
\end{center}

Proposer 1 initiates the \readPhase{} in ballot 0 for value $x$. It selects
$\set{A, B, D, E}$ as its \readQuorum{} and $\set{B, C}$ as its intent. It
sends prepare messages to the \readQuorum{}, and the \readQuorum{} replies.
Proposer 1 doesn't receive any intents, so it does not expand its
\readQuorum{}. It also learns that no value has been chosen yet, so it proposes
value $x$ to $B$ and $C$. Both accept the value.

\begin{center}
  \begin{tikzpicture}[xscale=1]
    \nodes
    \node[right=0cm of A] {\scriptsize $0  , -1 , \bot , \set{0:\set{B, C}}$};
    \node[right=0cm of B] {\scriptsize $0  , 0  , x    , \set{0:\set{B, C}}$};
    \node[right=0cm of C] {\scriptsize $0  , 0  , x    , \emptyset$};
    \node[right=0cm of D] {\scriptsize $0  , -1 , \bot , \set{0:\set{B, C}}$};
    \node[right=0cm of E] {\scriptsize $0  , -1 , \bot , \set{0:\set{B, C}}$};
    \node[right=0cm of F] {\scriptsize $-1 , -1 , \bot , \emptyset$};
    \node[right=0cm of G] {\scriptsize $-1 , -1 , \bot , \emptyset$};
    \node[right=0cm of H] {\scriptsize $-1 , -1 , \bot , \emptyset$};
    \node[right=0cm of I] {\scriptsize $-1 , -1 , \bot , \emptyset$};
  \end{tikzpicture}
\end{center}

Next, proposer 2 initiates the \readPhase{} in ballot 1 for value $y$. It
selects $\set{E, F, H, I}$ as its \readQuorum{} and $\set{G, H}$ as its intent.
It sends prepare messages to the \readQuorum{}, and the \readQuorum{} replies.
Proposer 2 receives the intent $\set{B, C}$ in ballot 0 from $E$, so it expands
its \readQuorum{} and sends a prepare message to $C$. Proposer 2 learns that
value $x$ was chosen in ballot $0$, so it ditches $y$ and proposes $x$ to $G$
and $H$. $G$ accepts, but the propose message to $H$ is dropped.

\begin{center}
  \begin{tikzpicture}[xscale=1]
    \nodes
    \node[right=0cm of A] {\scriptsize $0 , -1 , \bot , \set{0:\set{B, C}}$};
    \node[right=0cm of B] {\scriptsize $0 , 0  , x    , \set{0:\set{B, C}}$};
    \node[right=0cm of C] {\scriptsize $0 , 0  , x    , \set{1:\set{G, H}}$};
    \node[right=0cm of D] {\scriptsize $0 , -1 , \bot , \set{0:\set{B, C}}$};
    \node[right=0cm of E] {\scriptsize $1 , -1 , \bot , \set{0:\set{B, C}, 1:\set{G, H}}$};
    \node[right=0cm of F] {\scriptsize $1 , -1 , \bot , \set{1:\set{G, H}}$};
    \node[right=0cm of G] {\scriptsize $1 ,  1 , x, \emptyset$};
    \node[right=0cm of H] {\scriptsize $1 , -1 , \bot , \set{1:\set{G, H}}$};
    \node[right=0cm of I] {\scriptsize $1 , -1 , \bot , \set{1:\set{G, H}}$};
  \end{tikzpicture}
\end{center}

Next, garbage collection is run. The garbage collector contacts $G$ and sees
that it has accepted a value in ballot 1. It informs all the nodes to discard
intents in ballots less than 1.

\begin{center}
  \begin{tikzpicture}[xscale=1]
    \nodes
    \node[right=0cm of A] {\scriptsize $0 , -1 , \bot , \emptyset{}$};
    \node[right=0cm of B] {\scriptsize $0 , 0  , x    , \emptyset{}$};
    \node[right=0cm of C] {\scriptsize $0 , 0  , x    , \set{1:\set{G, H}}$};
    \node[right=0cm of D] {\scriptsize $0 , -1 , \bot , \emptyset{}$};
    \node[right=0cm of E] {\scriptsize $1 , -1 , \bot , \set{1:\set{G, H}}$};
    \node[right=0cm of F] {\scriptsize $1 , -1 , \bot , \set{1:\set{G, H}}$};
    \node[right=0cm of G] {\scriptsize $1 , 1 , x, \emptyset$};
    \node[right=0cm of H] {\scriptsize $1 , -1 , \bot , \set{1:\set{G, H}}$};
    \node[right=0cm of I] {\scriptsize $1 , -1 , \bot , \set{1:\set{G, H}}$};
  \end{tikzpicture}
\end{center}

Next, proposer 3 initiates the \readPhase{} in ballot 2 for value $z$ It
selects $\set{D, E, H, I}$ as its \readQuorum{} and $\set{E, F}$ as its intent.
It sends prepare messages to the \readQuorum{}, and the \readQuorum{} replies.
Proposer 3 receives intent $\set{G, H}$ in ballot 1, but has already included
$H$ in its \readQuorum{}, so it does not send any additional prepares. It
learns that no value has been chosen (this is a bug, $x$ was chosen), so it
proposes value $z$ to $E$ and $G$. Both accept the value, and $z$ is chosen.
This is a bug since $x$ was already chosen.

\begin{center}
  \begin{tikzpicture}[xscale=0.9]
    \nodes
    \node[right=0cm of A] {\scriptsize $0 , -1 , \bot , \emptyset{}$};
    \node[right=0cm of B] {\scriptsize $0 , 0  , x    , \emptyset{}$};
    \node[right=0cm of C] {\scriptsize $0 , 0  , x    , \set{1:\set{G, H}}$};
    \node[right=0cm of D] {\scriptsize $2 , -1 , \bot , \emptyset{2:\set{E, F}}$};
    \node[right=0cm of E] {\scriptsize $2 , 2  , z    , \set{1:\set{G, H}, 2:\set{E, F}}$};
    \node[right=0cm of F] {\scriptsize $2 , 2  , z    , \set{1:\set{G, H}}$};
    \node[right=0cm of G] {\scriptsize $1 , 1  , x    , \emptyset$};
    \node[right=0cm of H] {\scriptsize $2 , -1 , \bot , \set{1:\set{G, H}, 2: \set{E, F}}$};
    \node[right=0cm of I] {\scriptsize $2 , -1 , \bot , \set{1:\set{G, H}, 2: \set{E, F}}$};
  \end{tikzpicture}
\end{center}
\end{techreport}

\begin{techreport}

  \textbf{Cheap Paxos.}
  Cheap Paxos~\cite{lamport2004cheap} is a MultiPaxos variant that consists of
  a fixed set of $f+1$ main acceptors and $f$ auxiliary acceptors. During
  failure-free execution (the normal case) only the main acceptors are contacted.
  The auxiliary acceptors perform MultiPaxos' horizontal reconfiguration protocol
  to replace failed main acceptors.
  As with Fast Paxos, we can deploy Matchmaker MultiPaxos with only $f+1$
  acceptors, $f$ fewer than Cheap Paxos. Matchmaker Paxos does require $2f+1$
  matchmakers, but matchmakers do not act as acceptors and have to process only a
  single message (i.e.\ a \textsc{MatchA} message) to perform a reconfiguration.
\end{techreport}

\subsection{Generality}
First, we elaborate on MultiPaxos' horizontal reconfiguration. To reconfigure
from a set of nodes $N$ to a new set of nodes $N'$, the MultiPaxos leader gets
the value $N'$ chosen in the log at some index $i$. All commands in the log
starting at position $i + \alpha$ are chosen using the nodes in $N'$ instead of
the nodes in $N$, where $\alpha$ is some configurable parameter. Note that the
MultiPaxos leader can process at most $\alpha$ unchosen commands at a time.

%

\iftoggle{techreportenabled}{%
  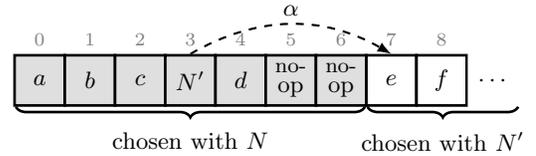
\begin{figure}[ht]
  \centering
  \tikzstyle{logentry}=[draw, line width=1pt, minimum width=19pt,
                        minimum height=19pt, align=center]
  \tikzstyle{lognum}=[gray, font=\scriptsize]
  \tikzstyle{oldconfig}=[fill=gray!25]
  \tikzstyle{newconfig}=[]
  \tikzstyle{brace}=[decorate, decoration={brace, mirror, amplitude=4pt},
                     line width=1pt]
  \tikzstyle{chosenlabel}=[align=center, below=6pt]
  \tikzstyle{pointer}=[draw, thick, -latex, dashed]
  \begin{tikzpicture}
    \node[logentry, label={[lognum]0}, oldconfig] (l0) at (0, 0) {$a$};
    \node[logentry, label={[lognum]1}, oldconfig, right=-1pt of l0] (l1) {$b$};
    \node[logentry, label={[lognum]2}, oldconfig, right=-1pt of l1] (l2) {$c$};
    \node[logentry, label={[lognum]3}, oldconfig, right=-1pt of l2] (l3) {$N'$};
    \node[logentry, label={[lognum]4}, oldconfig, right=-1pt of l3] (l4) {$d$};
    \node[logentry, label={[lognum]5}, oldconfig, right=-1pt of l4] (l5) {\small no-\\[-3pt]op};
    \node[logentry, label={[lognum]6}, oldconfig, right=-1pt of l5] (l6) {\small no-\\[-3pt]op};
    \node[logentry, label={[lognum]7}, newconfig, right=-1pt of l6] (l7) {$e$};
    \node[logentry, label={[lognum]8}, newconfig, right=-1pt of l7] (l8) {$f$};
    \node[logentry, draw=white, right=0 of l8] (dots) {$\cdots$};
    \draw[pointer, bend left] (l3.north) to node[above]{$\alpha$} (l7.north);

    \draw[brace] ([xshift=1pt] l0.south west) to %
                 node[chosenlabel]{chosen with $N$} %
                 ([xshift=-1pt] l6.south east);
    \draw[brace] ([xshift=1pt] l7.south west) to %
                 node[chosenlabel]{chosen with $N'$} %
                 ([xshift=-1pt] dots.south east);
  \end{tikzpicture}
  \caption{A MultiPaxos log during reconfiguration ($\alpha = 4$).}%
  \figlabel{HorizontalMultiPaxosReconfiguration}
\end{figure}
}{}

MultiPaxos' horizontal reconfiguration protocol is simple and has been proven
correct. Unfortunately, it is fundamentally incompatible with replication
protocols that do not have a log. Moreover, researchers are finding that
avoiding a log can often be advantageous~\cite{lamport2005generalized,
moraru2013there, whittaker2019bipartisan, arun2017speeding,
rystsov2018caspaxos, zhang2018building, szekeres20meerkat}.

\begin{techreport}

  For example, protocols like Generalized Paxos~\cite{lamport2005generalized},
  EPaxos~\cite{moraru2013there}, Janus~\cite{mu2016consolidating},
  BPaxos~\cite{whittaker2019bipartisan}, and Caesar~\cite{arun2017speeding}
  arrange commands in a partially ordered graph instead of a totally ordered log
  to exploit commutativity between commands. CASPaxos~\cite{rystsov2018caspaxos}
  maintains a single value, instead of a log or graph, for simplicity. Databases
  like TAPIR~\cite{zhang2018building} avoid ordering transactions in a log for
  improved performance, and databases like Meerkat~\cite{szekeres20meerkat} do
  the same to improve scalability.
\end{techreport}

Because these protocols do not have logs, they cannot use MultiPaxos'
horizontal reconfiguration protocol. However, while none of the protocols have
logs, \emph{all} of them have rounds. This means that the protocols can either
use Matchmaker Paxos directly, or at least borrow ideas from Matchmaker Paxos
for reconfiguration.
\begin{revisions}
  BPaxos~\cite{whittaker2020matchmaker}, for example, is an
  EPaxos~\cite{moraru2013there} variant that partially orders commands into a
  graph. BPaxos is a modular protocol that uses Paxos as a black box
  subroutine. Due to this modularity, we can directly replace Paxos with
  Matchmaker Paxos to support reconfiguration. The same idea can also be
  applied to EPaxos.
\end{revisions}\ %
CASPaxos~\cite{rystsov2018caspaxos} is almost identical to Paxos and can be
extended to Matchmaker CASPaxos in the same way we extended Paxos to Matchmaker
Paxos. These are two simple examples, and we don't claim that extending
Matchmaker Paxos to some of the other more complicated protocols is always
easy. But, the universality of rounds makes Matchmaker Paxos an attractive
foundation on top of which other non-log based protocols can build their own
reconfiguration protocols.

}
{\section{Evaluation}\seclabel{Evaluation}
We now evaluate Matchmaker MultiPaxos. Refer to our technical
report~\cite{whittaker2020matchmaker} for further evaluation.
Our implementation of Matchmaker MultiPaxos is available at
\begin{revisions}
  \url{github.com/mwhittaker/frankenpaxos}.
\end{revisions}\ %
Matchmaker MultiPaxos is implemented in Scala using the Netty networking
library. We deployed Matchmaker MultiPaxos on m5.xlarge AWS EC2 instances
within a single availability zone. For a given $f$, we deployed $f+1$
proposers, $2f+1$ acceptors, $2f+1$ matchmakers, and $2f+1$ replicas. For
simplicity, every node is deployed on its own machine, but in practice, nodes
can be physically co-located.
\begin{revisions}
  In particular, any two logical roles can be placed on the same machine, so
  long as the two roles are not the same. For example, we can co-locate a
  leader, an acceptor, a replica, and a matchmaker, but we can't co-locate two
  acceptors (without reducing the fault tolerance of the system).
\end{revisions}\ %
All of our results hold in a co-located deployment as well.
For simplicity, we deploy Matchmaker MultiPaxos with a trivial no-op state
machine in which every state machine command is a one byte no-op.

%
%
%

\subsection{Reconfiguration}\seclabel{EvalReconfiguration}
\textbf{Experiment Description.}
For each $f$, we run a benchmark with 1, 4, and 8 clients.
Every client repeatedly proposes a state machine command, waits to receive a
response, and then immediately proposes another command. Every benchmark runs
for 35 seconds. During the first 10 seconds, we perform no reconfigurations.
From 10 seconds to 20 seconds, the leader reconfigures the set of acceptors
once every second. In practice, we would reconfigure much less often.
For each of the ten reconfigurations, the leader selects a random set of $2f+1$
acceptors from a pool of $2\times(2f+1)$ acceptors. At 25 seconds, we fail one
of the acceptors. 5 seconds later, the leader performs a reconfiguration to
replace the failed acceptor. The delay of 5 seconds is completely arbitrary.
The leader can reconfigure sooner if desired.


\begin{revisions}
  We also perform this experiment with an implementation of MultiPaxos with
  horizontal reconfiguration.
  %
  We set $\alpha$ to $8$. Because $\alpha$ is equal to the number of clients,
  MultiPaxos never stalls because of an insufficiently large $\alpha$.
\end{revisions}

\begin{figure}[t]
  \centering
  \includegraphics[width=\columnwidth]{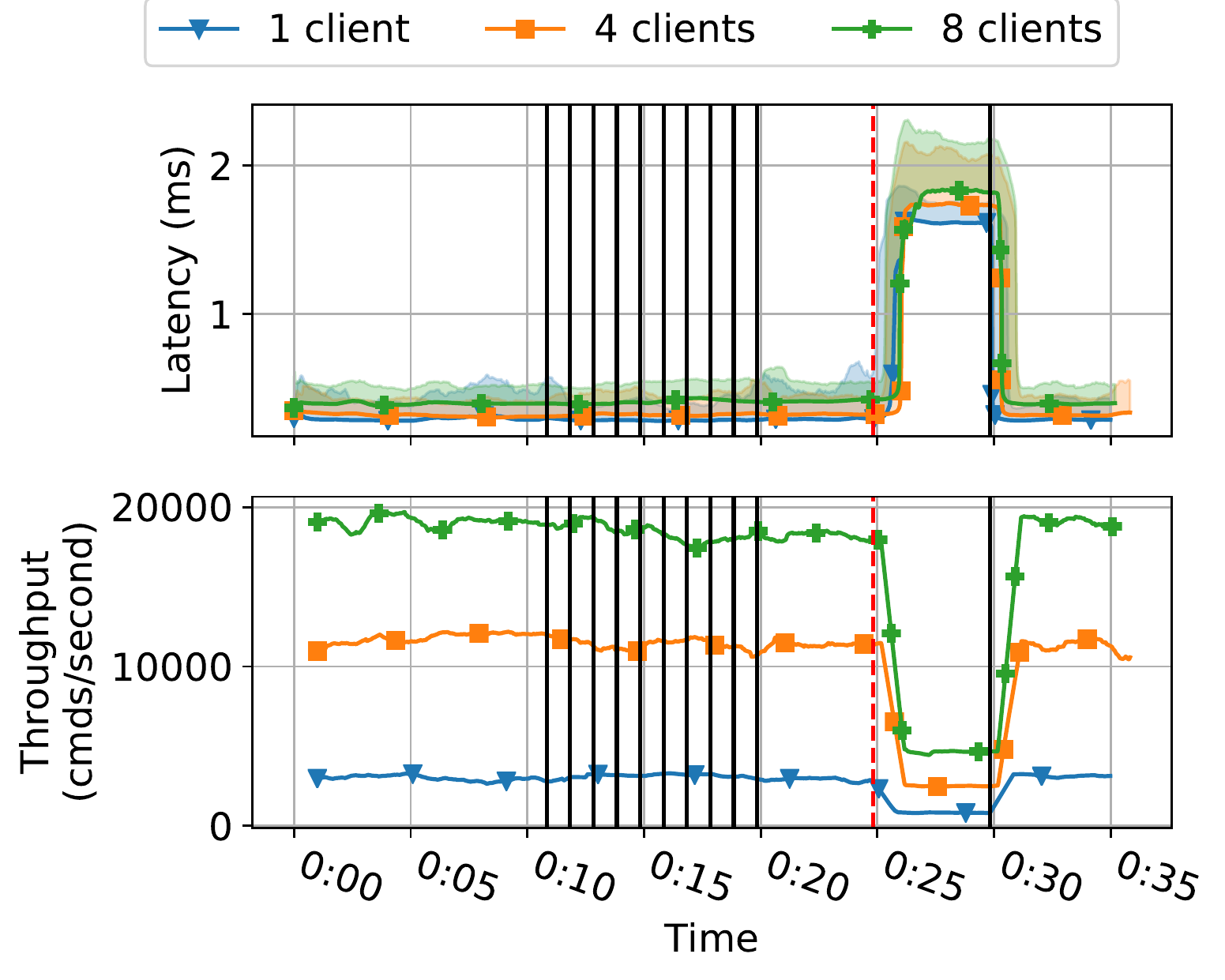}
  \caption{%
    Matchmaker MultiPaxos' latency and throughput ($f=1$).
    \markrevisions{
      Median latency is shown using solid lines, while the 95\% latency is
      shown as a shaded region above the median latency.
    }%
    The vertical black lines show reconfigurations. The vertical dashed red
    line shows an acceptor failure.
  }\figlabel{LeaderReconfigurationF1}
\end{figure}

\begin{figure}[t]
  \centering
  \includegraphics[width=\columnwidth]{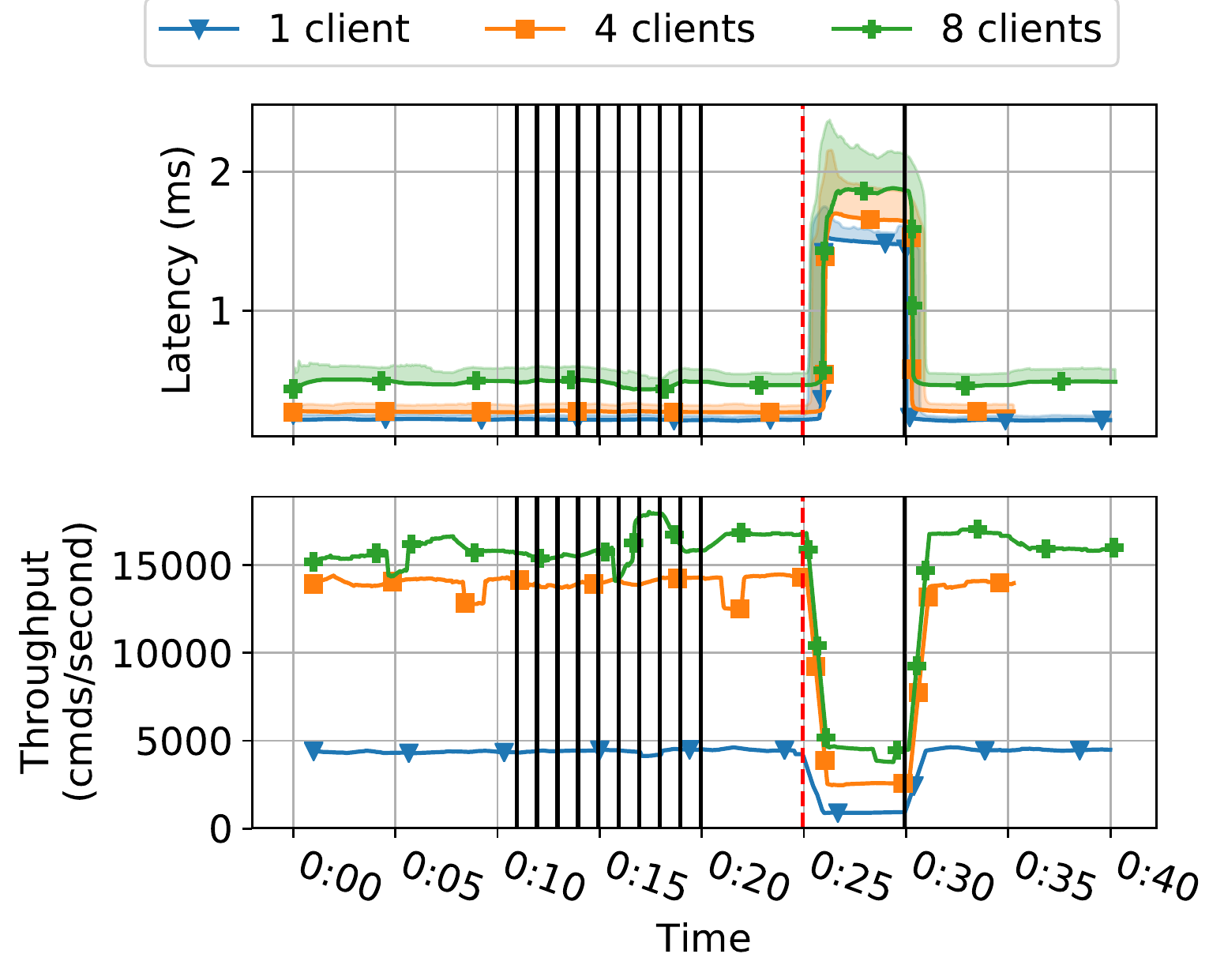}
  \caption{%
    \markrevisions{%
      The latency and throughput of MultiPaxos with horizontal reconfiguration
      ($f=1$).
    }
  }\figlabel{HorizontalLeaderReconfiguration}
\end{figure}

\iftoggle{techreportenabled}{%
  \begin{figure}[ht]
    \centering
    \includegraphics[width=\columnwidth]{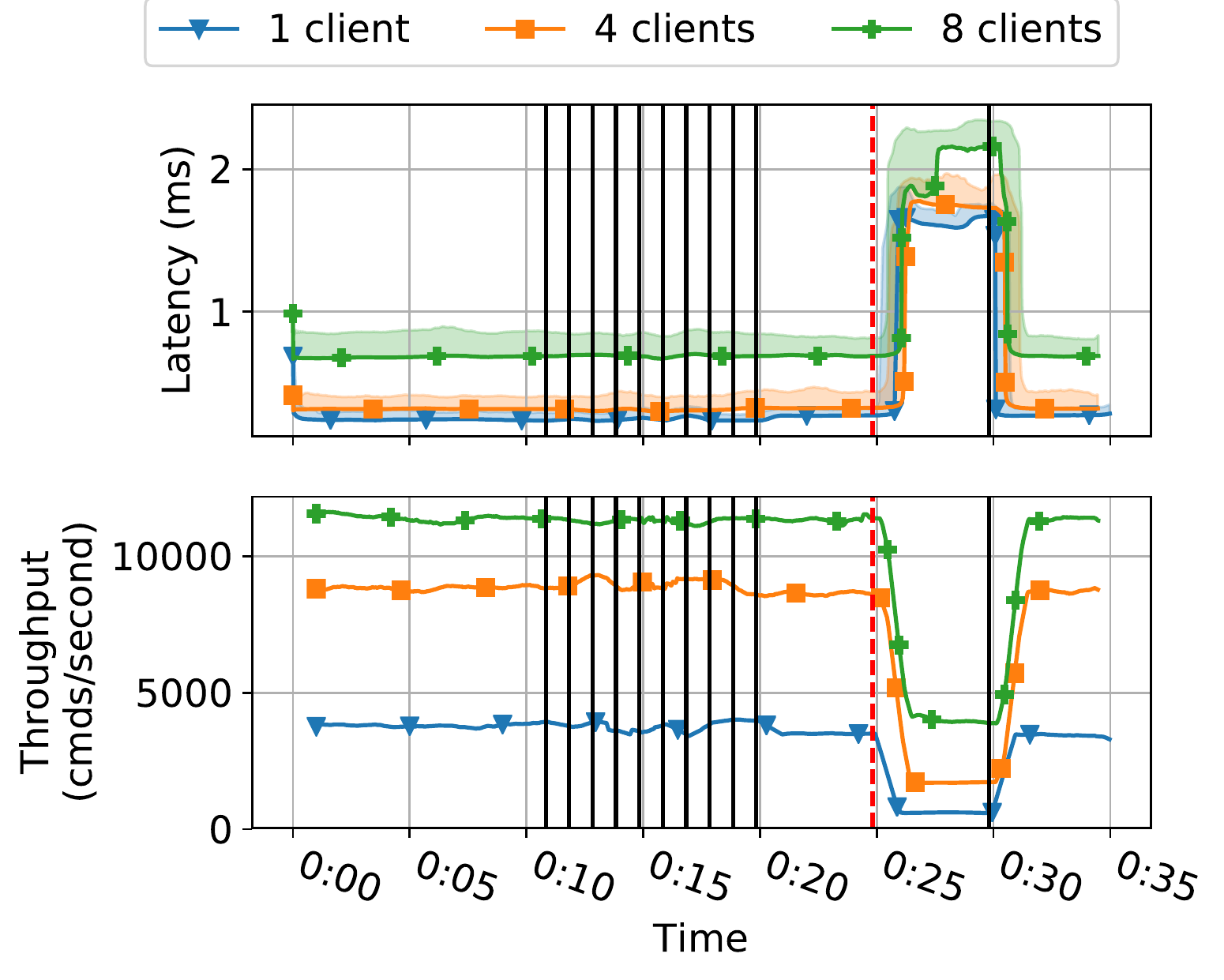}
    \caption{%
      Matchmaker MultiPaxos' latency and throughput ($f=2$). The vertical black
      lines show reconfigurations. The vertical dashed red line shows an
      acceptor failure.
    }\figlabel{LeaderReconfigurationF2}
  \end{figure}
}{}

\iftoggle{techreportenabled}{%
  \begin{figure}[ht]
    \centering
    \includegraphics[width=\columnwidth]{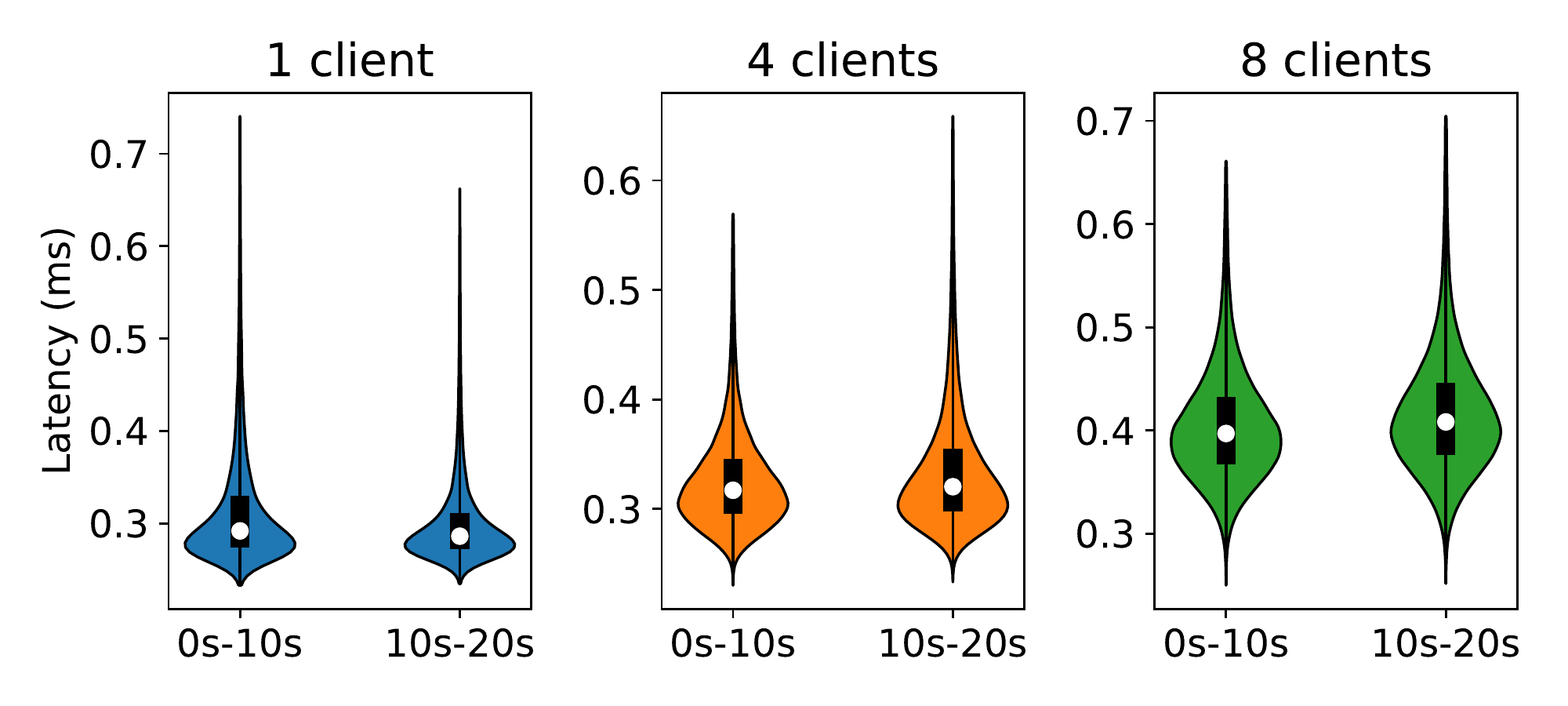}
    \includegraphics[width=\columnwidth]{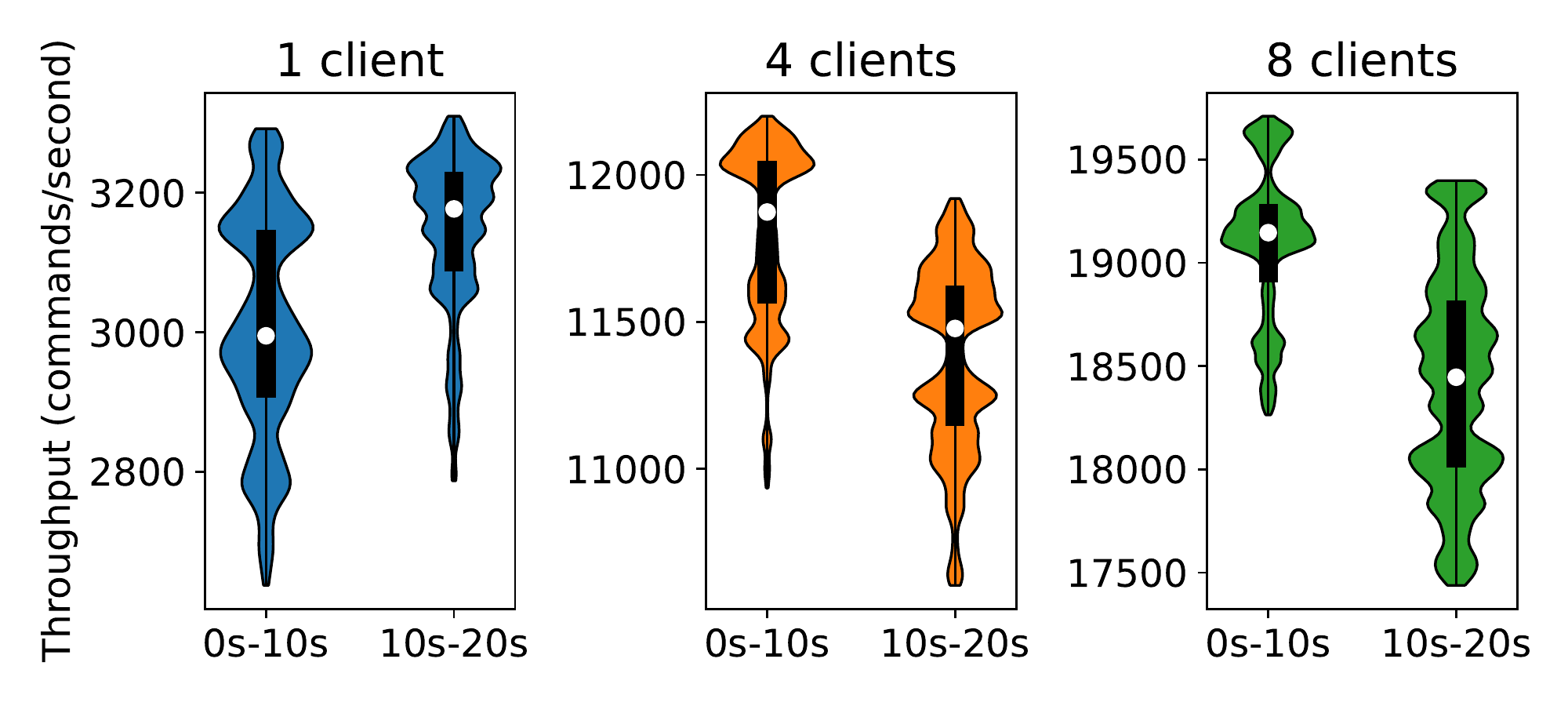}
    \caption{%
      Violin plots of \figref{LeaderReconfigurationF1} latency and throughput
      during the first 10 seconds and between 10 and 20 seconds.
    }\figlabel{Violin}
  \end{figure}
}{}

\iftoggle{techreportenabled}{%
  \begin{figure}[ht]
    \centering
    \includegraphics[width=\columnwidth]{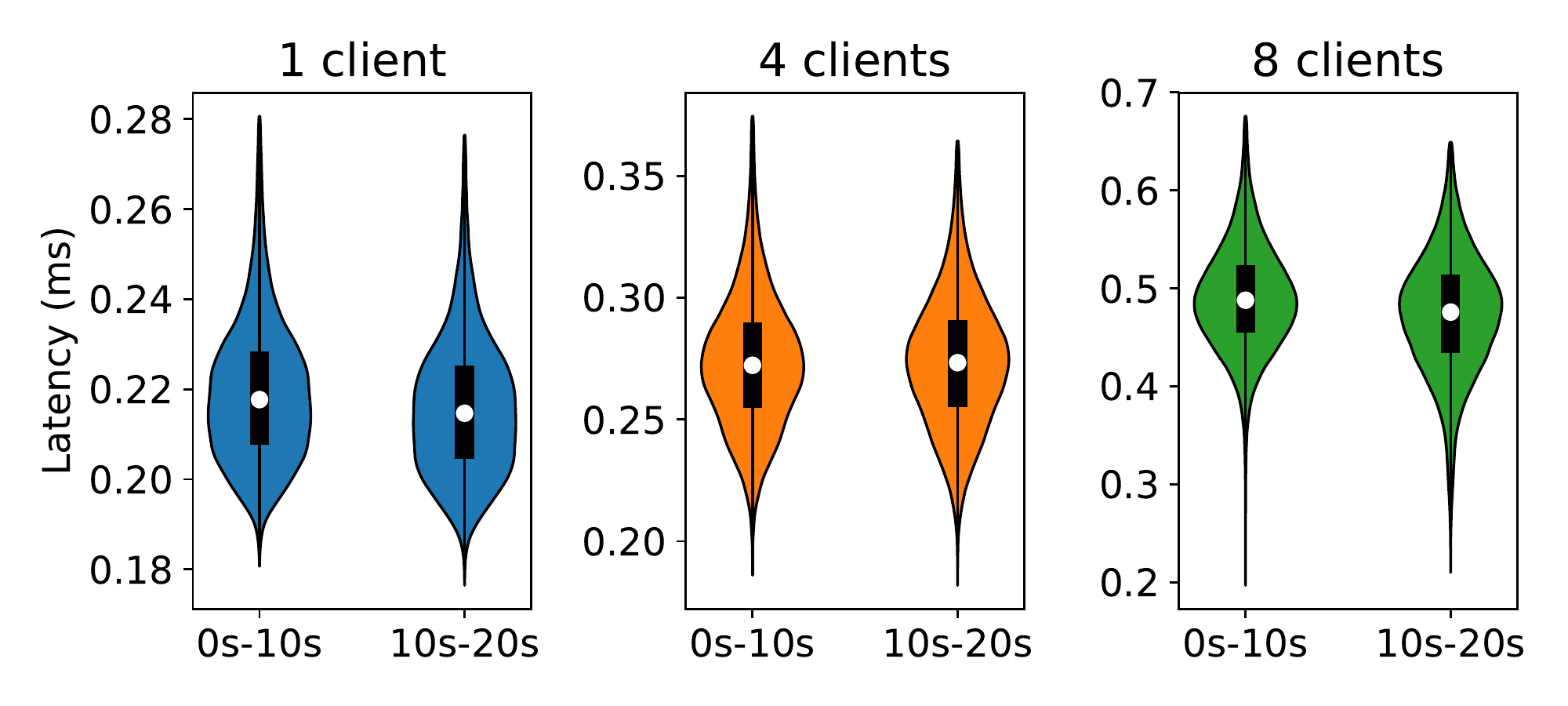}
    \includegraphics[width=\columnwidth]{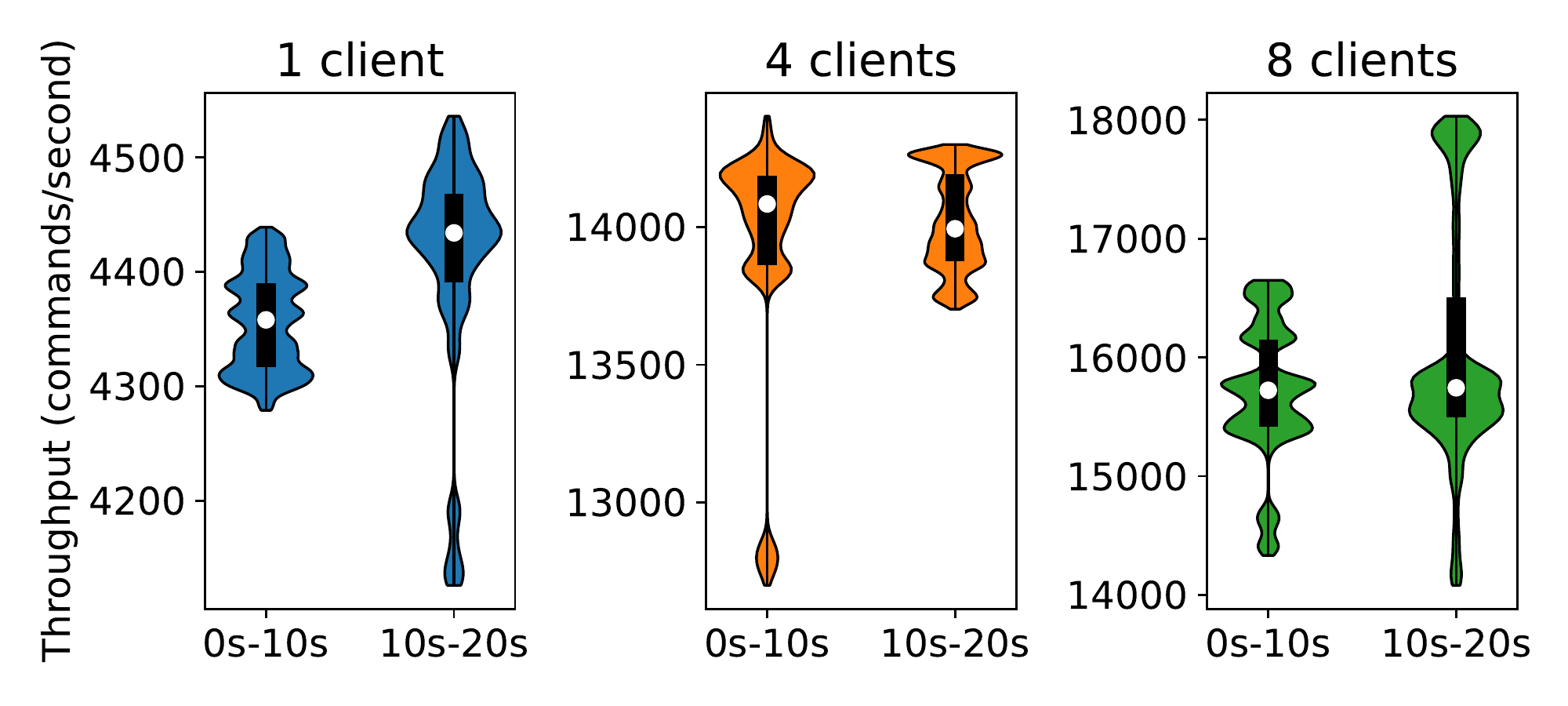}
    \caption{%
      \markrevisions{%
        Violin plots of \figref{HorizontalLeaderReconfiguration} latency and
        throughput during the first 10 seconds and between 10 and 20 seconds.
      }
    }\figlabel{Violin}
  \end{figure}
}{}

\textbf{Results.}
The latency and throughput of Matchmaker MultiPaxos for $f=1$ is shown in
\figref{LeaderReconfigurationF1}.
\iftoggle{techreportenabled}{}{%
  For space reasons, we report all of our results only for $f=1$.
  \begin{revisions}
    The results for larger values of $f$ are similar and shown in our technical
    report~\cite{whittaker2020matchmaker}. Note that Matchmaker MultiPaxos'
    throughput decreases and its latency increases as we increase $f$. The
    overall trends, however, are the same.
  \end{revisions}\ %
}
Throughput and latency are both computed using sliding one second windows.
Median latency is shown using solid lines, while the 95\% latency is shown as a
shaded region above the median latency. The black vertical lines denote
reconfigurations, and the red dashed vertical line denotes the acceptor
failure.

The medians, interquartile ranges (IQR), and standard deviations of the latency
and throughput (a) during the first 10 seconds and (b) between 10 and 20
seconds are shown in \tabref{LeaderReconfigurationValues}.
\iftoggle{techreportenabled}{%
  \figref{Violin} includes violin plots of the same data. The white circles
  show the median values, while the thick black rectangles show the \nth{25}
  and \nth{75} percentiles.
}{}
For latency, reconfiguration has little to no impact (roughly 2\% changes) on
the medians, IQRs, or standard deviations. The one exception is that the 8
client standard deviation is significantly larger. This is due to a small
number of outliers. Reconfiguration has little impact on median throughput,
with all differences being statistically insignificant. The IQRs and standard
deviations \markrevisions{sometimes increase and sometimes decrease}. The IQR
is always less than 1\% of the median throughput, and the standard deviation is
always less than 4\%.

For every reconfiguration, the new acceptors become active within a
millisecond. The old acceptors are garbage collected within five milliseconds.
\begin{revisions}
  This means that only one configuration is ever returned by the matchmakers.
\end{revisions}\ %
We implement Matchmaker MultiPaxos with an optimization called
thriftiness~\cite{moraru2013there}---where \textsc{Phase2A} messages are sent
to a randomly selected Phase 2 quorum---so the throughput and latency
expectedly degrade after we fail an acceptor. After we replace the failed
acceptor, throughput and latency return to normal within two seconds.

{\begin{table}[t]
  \centering
  \caption{%
    \figref{LeaderReconfigurationF1} median, interquartile range, and standard
    deviation of latency and throughput.
  }\tablabel{LeaderReconfigurationValues}
  Latency (ms)
  \begin{tabular}{l@{\hskip 4pt}c@{\hskip 4pt}cc@{\hskip 4pt}cc@{\hskip 4pt}c}
    \toprule
    & \multicolumn{2}{c}{1 Client}
    & \multicolumn{2}{c}{4 Clients}
    & \multicolumn{2}{c}{8 Clients} \\
    & 0s-10s & 10s-20s & 0s-10s & 10s-20s & 0s-10s & 10s-20s \\
    \cmidrule(lr){2-3}
    \cmidrule(lr){4-5}
    \cmidrule(lr){6-7}
    median & 0.292 & 0.287 & 0.317 & 0.321 & 0.398 & 0.410 \\
    IQR    & 0.040 & 0.026 & 0.029 & 0.036 & 0.036 & 0.039 \\
    stdev  & 0.114 & 0.085 & 0.076 & 0.081 & 0.089 & 0.305 \\
    \bottomrule
  \end{tabular}

  \vspace{12pt}
  Throughput (commands/second)
  \begin{tabular}{l@{\hskip 4pt}c@{\hskip 4pt}cc@{\hskip 4pt}cc@{\hskip 4pt}c}
    \toprule
    & \multicolumn{2}{c}{1 Client}
    & \multicolumn{2}{c}{4 Clients}
    & \multicolumn{2}{c}{8 Clients} \\
    & 0s-10s & 10s-20s & 0s-10s & 10s-20s & 0s-10s & 10s-20s \\
    \cmidrule(lr){2-3}
    \cmidrule(lr){4-5}
    \cmidrule(lr){6-7}
    median & 2,995 & 3,177 & 11,874 & 11,478 & 19,146 & 18,446 \\
    IQR    & 152   & 53    & 175    & 145    & 140    & 373 \\
    stdev  & 157   & 111   & 298    & 307    & 358    & 520 \\
    \bottomrule
  \end{tabular}
\end{table}
}

\begin{revisions}
  The latency and throughput of MultiPaxos for $f=1$ is shown in
  \figref{HorizontalLeaderReconfiguration}. As with Matchmaker MultiPaxos,
  MultiPaxos can perform a horizontal reconfiguration without any performance
  degradation.
\end{revisions}

\begin{techreportrevisions}

  Without thriftiness, a Matchmaker MultiPaxos leader sends \textsc{Phase2A}
  messages to \emph{all} of the acceptors, even though it only needs to hear
  back from a Phase 2 quorum of them. With thriftiness, on the other hand, a
  Matchmaker MultiPaxos leader only sends \textsc{Phase2A} messages to a Phase
  2 quorum of acceptors, rather than to all of them.

  Thriftiness is a trade-off between failure resilience and normal-case
  throughput. A non-thrifty leader has to send and receive more messages than a
  thrifty leader, so Matchmaker MultiPaxos' throughput is lower without
  thriftiness. However, because a thrifty leader only sends \textsc{Phase2A}
  messages to a Phase 2 quorum of acceptors, if any one of the acceptors in this
  Phase 2 quorum fails, then the leader is stuck. The leader has to wait for a
  timeout to expire before resending \textsc{Phase2A} messages to all of the
  acceptors.  This reduces throughput significantly. Thus, the throughput of
  thrifty Matchmaker MultiPaxos is higher in the normal case but lower when an
  acceptor fails.

  Note that the decision of whether to enable thriftiness is orthogonal to the
  design of Matchmaker MultiPaxos. Matchmaker MultiPaxos can be deployed with
  or without thriftiness.

  In \figref{NonThriftyLt}, we show the latency and throughput of Matchmaker
  MultiPaxos with and without thriftiness for various numbers of clients,
  without any failures. As expected, the peak throughput of Matchmaker
  MultiPaxos is higher when thriftiness is enabled. In
  \figref{NonThriftyLeaderReconfiguration}, we recreate
  \figref{LeaderReconfigurationF1}, but with thriftiness disabled. Again as
  expected, an acceptor failure has less of an effect on the latency and
  throughput of non-thrifty Matchmaker MultiPaxos.
\end{techreportrevisions}

\iftoggle{techreportenabled}{%
  \begin{figure}[ht]
    \centering
    \includegraphics[width=\columnwidth]{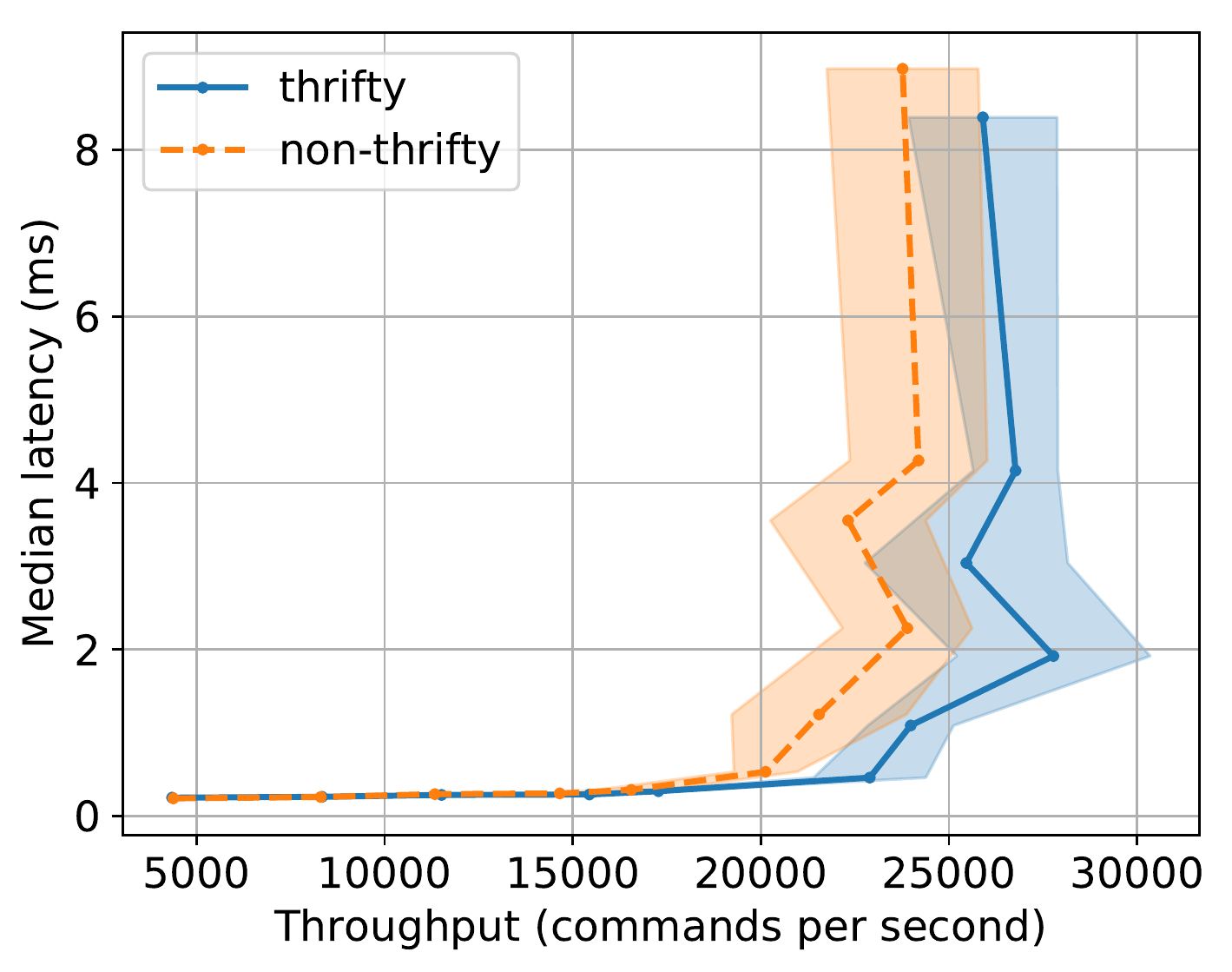}
    \caption{%
      \markrevisions{%
        Latency-throughput curves for Matchmaker MultiPaxos with and without
        thriftiness. The standard deviation of throughput measurements are shown
        as shaded regions around the mean.
      }
    }\figlabel{NonThriftyLt}
  \end{figure}
}{}

\iftoggle{techreportenabled}{%
  \begin{figure}[ht]
    \centering
    \includegraphics[width=\columnwidth]{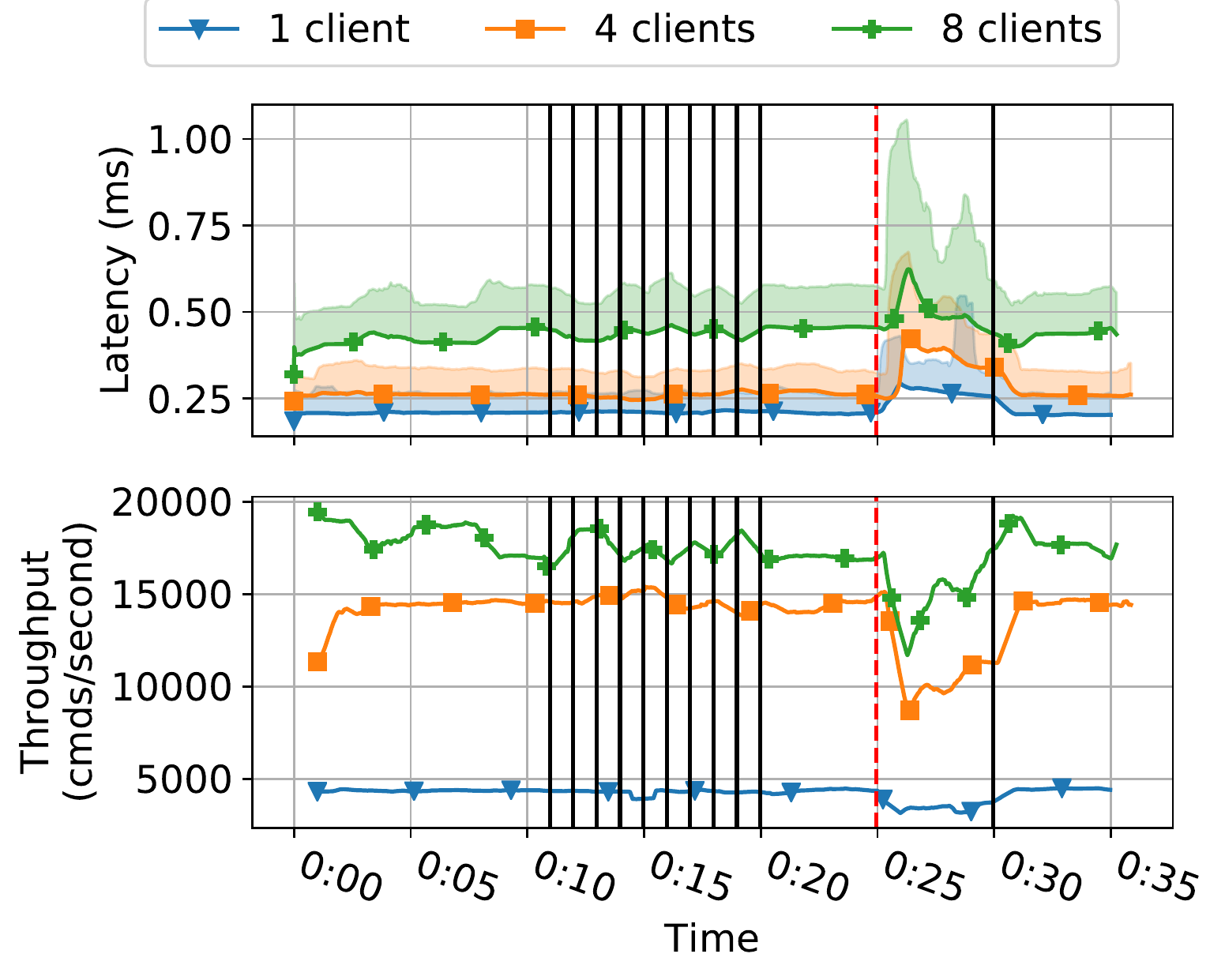}
    \caption{%
      \markrevisions{%
        A recreation of \figref{LeaderReconfigurationF1}, but without thriftiness.
      }
    }\figlabel{NonThriftyLeaderReconfiguration}
  \end{figure}
}{}

\begin{techreportrevisions}

  \figref{NonThriftyLt} shows that the peak throughput of Matchmaker MultiPaxos
  is higher than the throughput reported in \figref{LeaderReconfigurationF1}
  and requires more than eight clients. However, we intentionally perform our
  experiment with fewer than eight clients and below peak throughput. We do
  this to reduce the performance variations that naturally occur in our
  benchmarks.

  As Matchmaker MultiPaxos executes (even without any reconfigurations), the
  throughput and latency does not remain perfectly steady. For various reasons
  (e.g., JVM garbage collection, JVM just-in-time compilation, network jitter,
  operating system scheduling), the performance of the system varies over time.
  These performance variations increase as we add more clients. For example,
  more clients leads to higher throughput which triggers JVM garbage collection
  more often. More clients also leads to more congested network links which
  leads to more network jitter.

  All of our results still hold when we increase the number of clients and
  operate at peak load, but it becomes more difficult to tease apart which
  performance fluctuations occur naturally and which are caused by
  reconfiguration. \figref{MoreClientsLeaderReconfiguration}, for example,
  recreates \figref{LeaderReconfigurationF1} but with 100 clients. We see the
  same trends at higher throughput, but we also see more performance
  fluctuations, even when we are not performing any reconfigurations.
\end{techreportrevisions}

\iftoggle{techreportenabled}{%
  \begin{figure}[ht]
    \centering
    \includegraphics[width=\columnwidth]{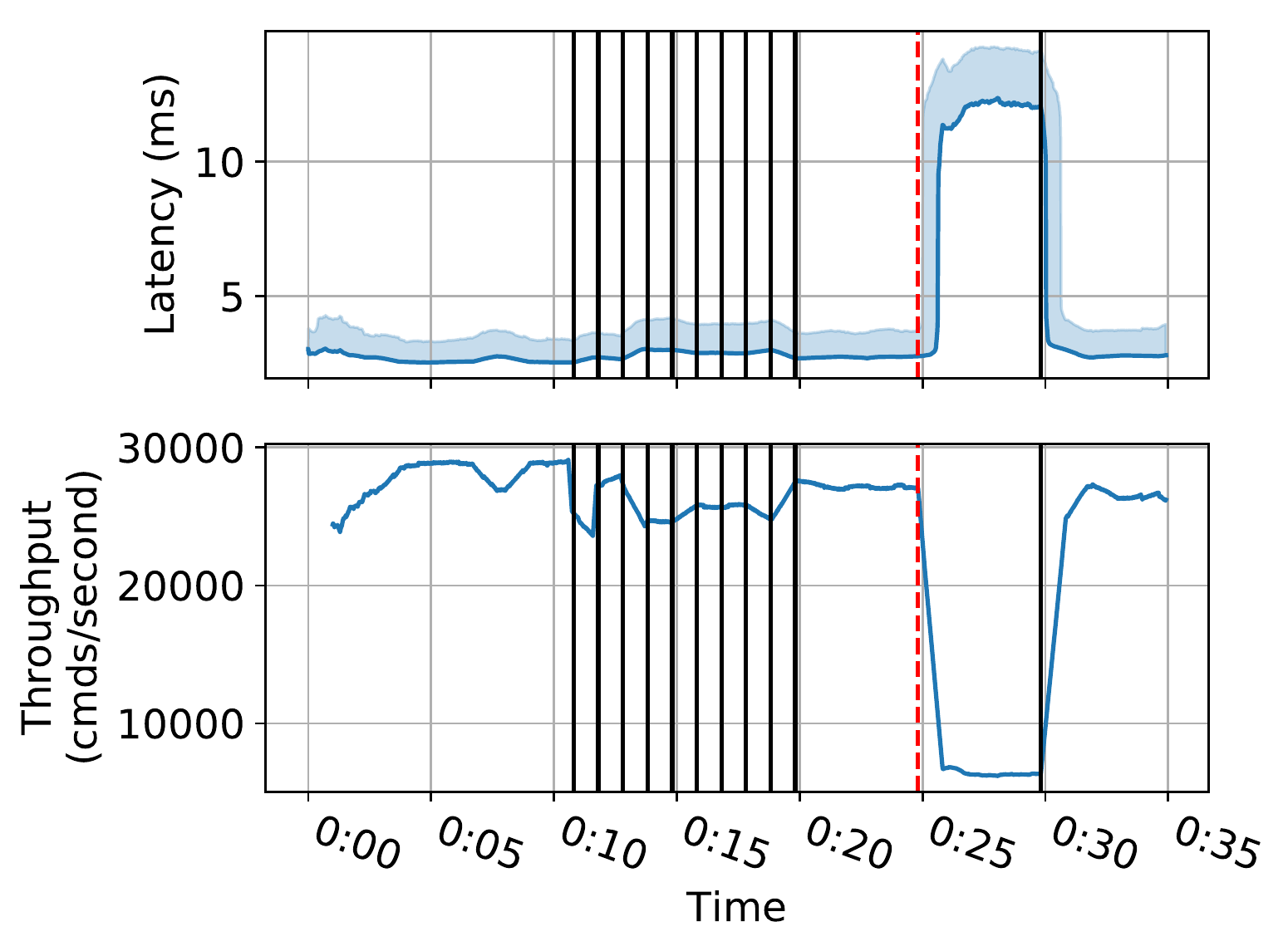}
    \caption{%
      \markrevisions{%
        A recreation of \figref{LeaderReconfigurationF1}, but with 100 clients.
      }
    }\figlabel{MoreClientsLeaderReconfiguration}
  \end{figure}
}{}

\begin{techreport}

  \textbf{Summary.}
  This experiment confirms that Matchmaker MultiPaxos's throughput and latency
  remain steady even during abnormally frequent reconfiguration. Moreover, it
  confirms that Matchmaker MultiPaxos can reconfigure to a new set of acceptors
  and retire the old set of acceptors on the order of milliseconds.
\end{techreport}

\begin{revisions}
  \subsection{Ablation Study}\seclabel{Ablation}

  \textbf{Experiment Description.}
  We now perform an ablation study to measure the effects of the three
  optimizations presented in \secref{MatchmakerPaxos}. We deploy Matchmaker
  MultiPaxos as above. Each benchmark runs for 20 seconds with eight clients.
  The leader reconfigures the set of acceptors five times throughout the course
  of the benchmark. Every benchmark is run with a different subset of the
  optimizations in \secref{MatchmakerPaxos} enabled.
  When garbage collection (Optimization 3) is disabled, no garbage collection
  is performed. When Phase 1 Bypassing (Optimization 2) is disabled, the leader
  does not process commands during Phase 1 of a reconfiguration. When Proactive
  Matchmaking (Optimization 3) is disabled, the leader does not process
  commands during the Matchmaking phase of a reconfiguration.
  To simulate running on a wide area network---a common
  scenario~\cite{corbett2013spanner, ailijiang2019wpaxos}---the acceptors and
  matchmakers delay their \textsc{Phase1B} and \textsc{MatchB} messages by 250
  milliseconds. To keep normal case throughout and latency consistent with the
  previous benchmarks, \textsc{Phase2B}s are not delayed.

  \begin{figure*}[htb]
    \centering
    \includegraphics[width=\textwidth]{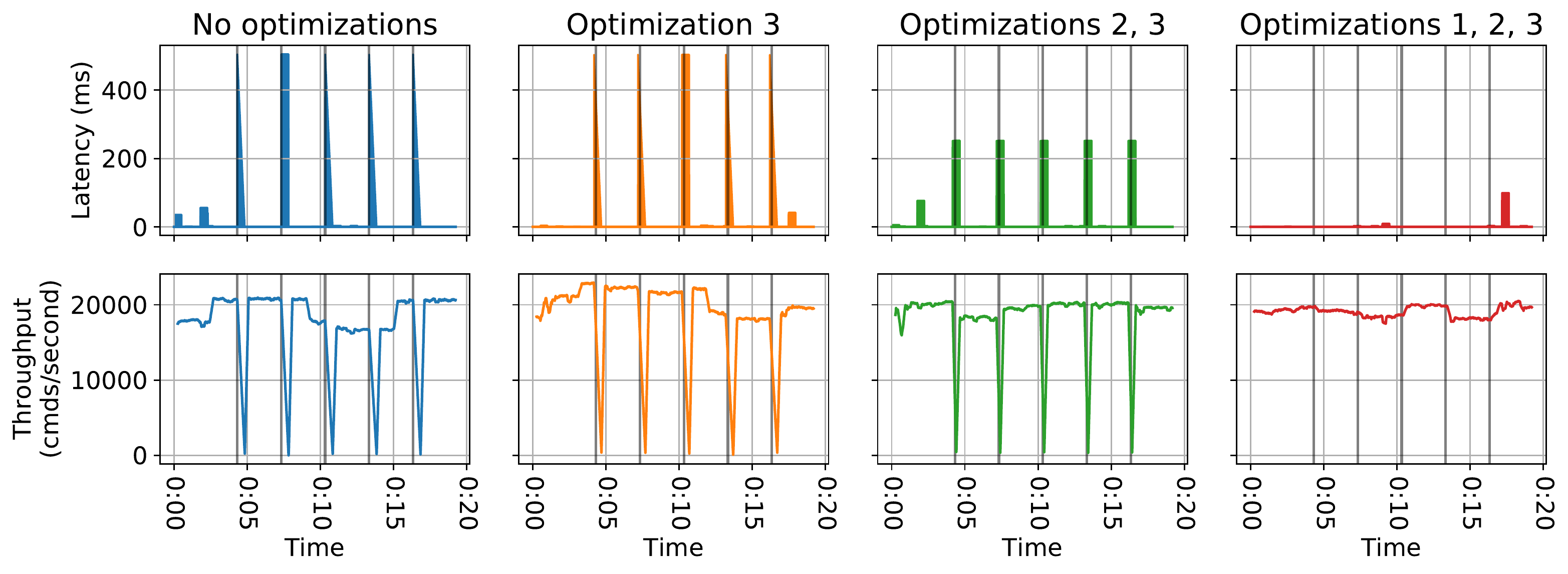}
    \caption{
      \markrevisions{
        The latency and throughput of Matchmaker MultiPaxos with various
        optimizations enabled.
      }
    }\figlabel{Ablation}
  \end{figure*}

  \textbf{Results.}
  The max latency (computed over 500 millisecond rolling windows) and
  throughput (250 millisecond windows) of Matchmaker MultiPaxos are shown in
  \figref{Ablation}. Without any optimizations, the latency of Matchmaker
  MultiPaxos expectedly spikes to 500 milliseconds after every reconfiguration,
  while the throughput drops to zero.
  %
  %
  %
  Enabling garbage collection has little effect on throughput or latency, but
  garbage collection is required to shut down old configurations, so we cannot
  disable it in practice.
  With garbage collection and Phase 1 Bypassing, the latency peaks only at 250
  milliseconds. The throughput drops to zero, but now only for 250
  milliseconds.
  With all three optimizations enabled, the performance of the protocol is
  steady.

  This ablation study serves three purposes.
  First, it highlights the importance of the three optimizations outlined in
  \secref{MatchmakerPaxos}.
  %
  Second, it highlights that Matchmaker MultiPaxos can mask the latency of a
  reconfiguration, even when the latency is large.
  %
  Third, the ablation study is indirectly a comparison with stop-the-world
  reconfiguration protocols like the one used by Viewstamped
  Replication~\cite{liskov2012viewstamped}. A stop-the-world reconfiguration
  protocol behaves like the unoptimized protocols in our ablation study. They
  stop processing commands entirely for some duration of time, so latency
  spikes and throughput drops to zero.
\end{revisions}

\subsection{Leader Failure}\seclabel{EvalLeaderFailure}
\textbf{Experiment Description.}
We deploy Matchmaker MultiPaxos exactly as before.
Now, each benchmark runs for 20 seconds. During the first 7 seconds, there are
no reconfigurations and no failures. At 7 seconds, we fail the leader. 5
seconds later, a new leader is elected and resumes normal operation. The 5
second delay is arbitrary; a new leader could be elected quicker if desired.

\begin{figure}[t]
  \centering
  \includegraphics[width=\columnwidth]{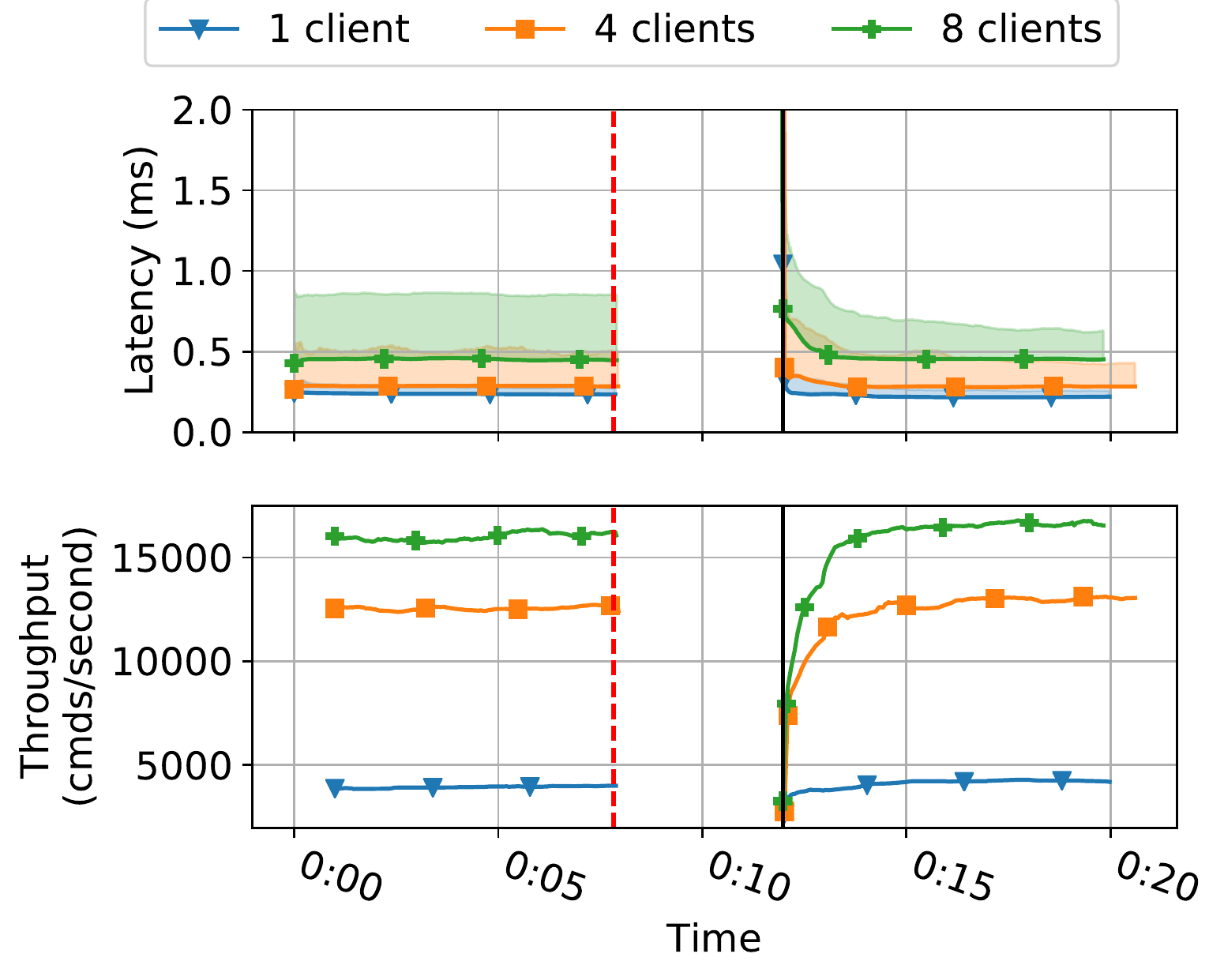}
  \caption{%
    Matchmaker MultiPaxos' latency and throughput ($f=1$).  The dashed red line
    denotes a leader failure.
  }\figlabel{LeaderFailureF1}
\end{figure}

\iftoggle{techreportenabled}{%
  \begin{figure}[ht]
    \centering
    \includegraphics[width=\columnwidth]{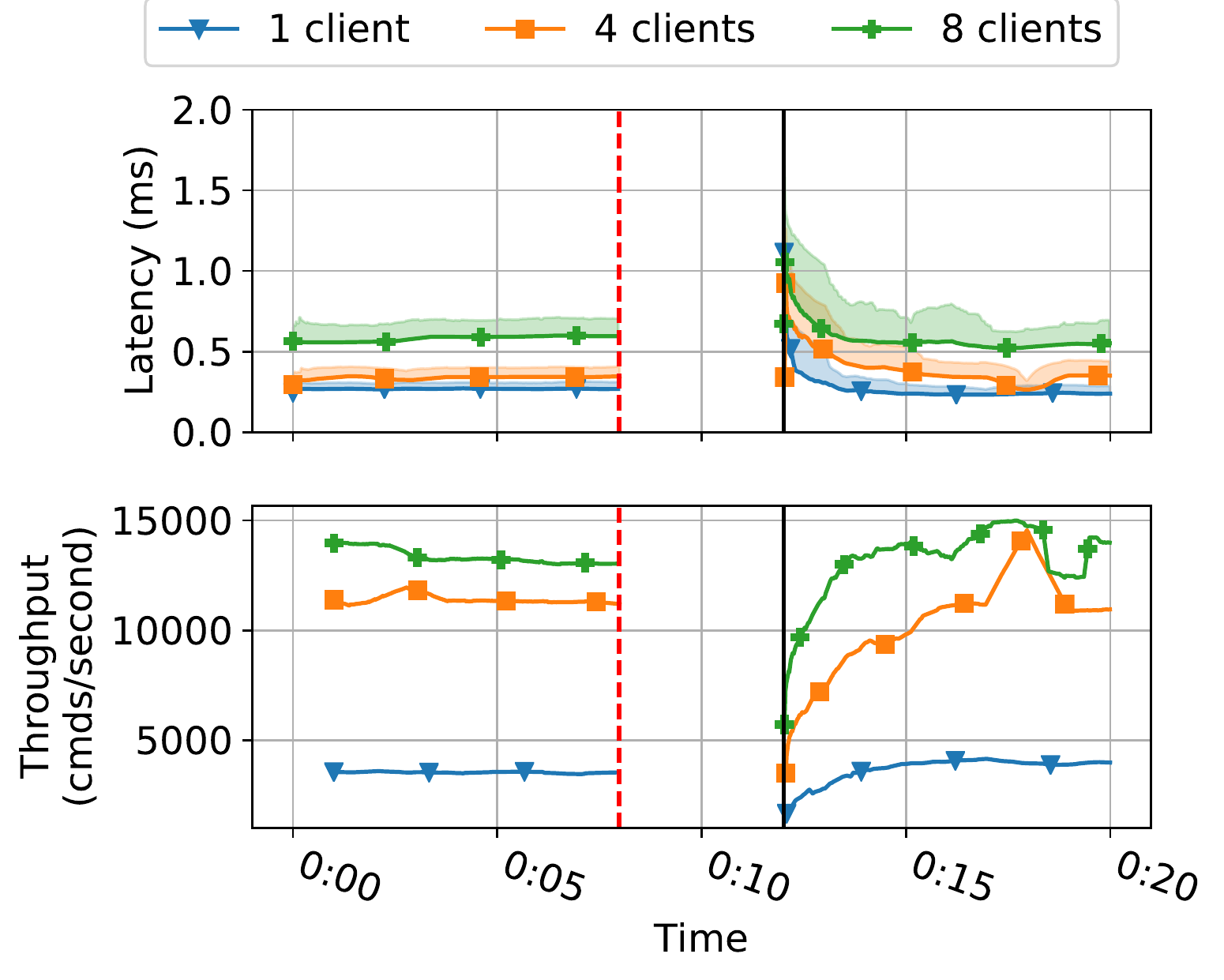}
    \caption{%
      \markrevisions{%
        The latency and throughput of Horizontal MultiPaxos with $f=1$.
      }
    }\figlabel{LeaderFailureF1}
  \end{figure}
}{}

\textbf{Results.}
The latency and throughput of the benchmarks are shown in
\figref{LeaderFailureF1}. During the first 7 seconds, throughput and latency
are both stable. When the leader fails, the throughput expectedly drops to
zero. The throughput and latency return to normal within two seconds after a
new leader is elected.
%
%

\begin{techreportrevisions}

  Matchmaker MultiPaxos can also tolerate many simultaneous failures. We repeat
  the experiment above but this time fail a leader, an acceptor, and a
  matchmaker at the same time, as shown in \figref{Chaos}. When the leader
  fails, the throughput expectedly drops to zero. Four seconds later, a new
  leader is elected and the protocol resumes execution, though the throughput
  is reduced because of the failed acceptor. Six seconds later, the leader
  reconfigures away from the failed acceptor and the throughput returns to
  normal. Five seconds later, the leader reconfigures away from the failed
  matchmaker, but since matchmaker reconfiguration is off the critical path, it
  has no effect on performance (this is discussed more in the next subsection).

  Note that the delays we use for reconfiguration are arbitrary. The leader,
  acceptor, and matchmaker can all be reconfigured simultaneously. We
  artificially stagger their reconfiguration to better see the effects of each.
\end{techreportrevisions}

\iftoggle{techreportenabled}{%
  \begin{figure}[ht]
    \centering
    \includegraphics[width=\columnwidth]{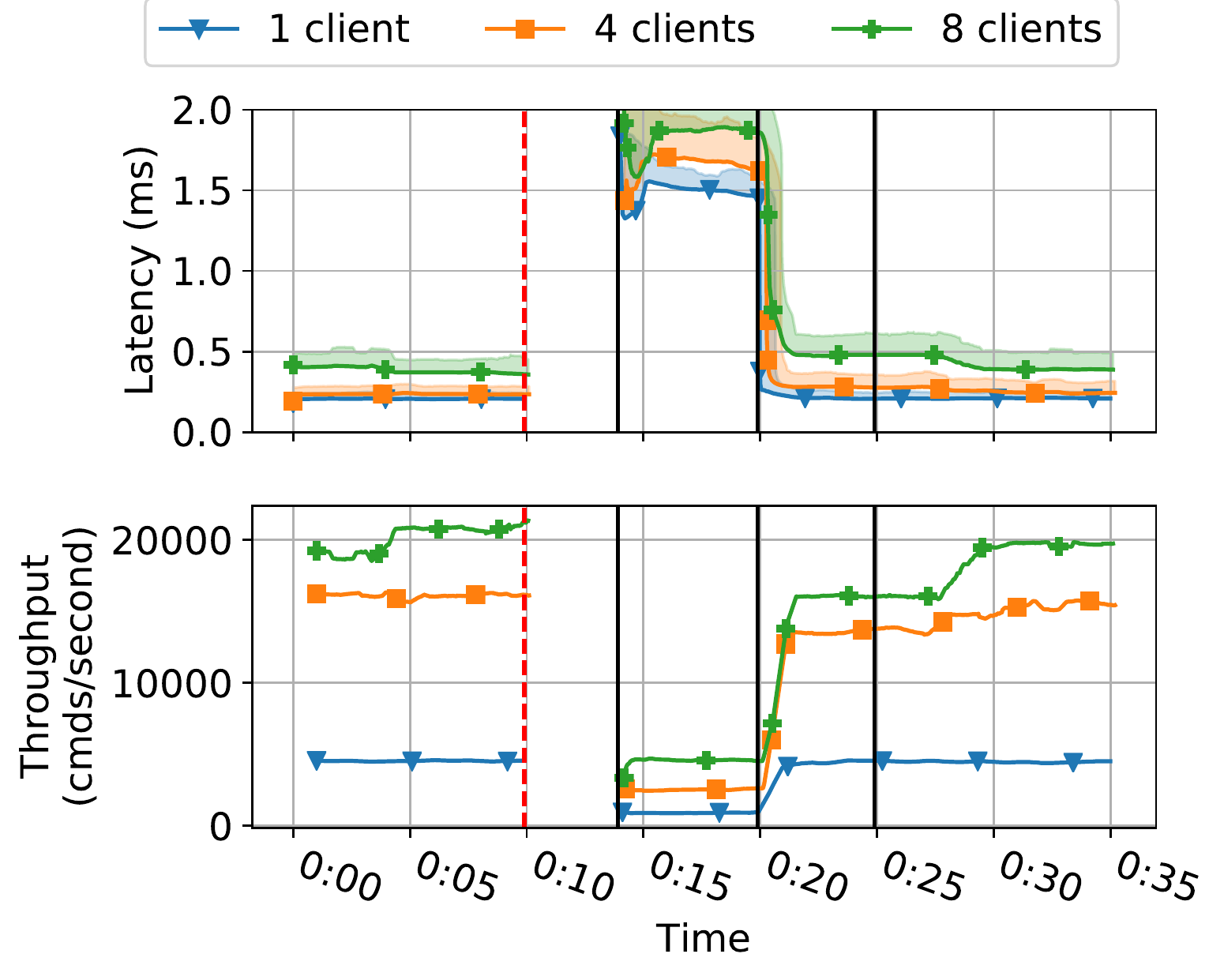}
    \caption{%
      \markrevisions{%
        The latency and throughput of Matchmaker MultiPaxos with $f=1$. The
        vertical dashed red line shows the failure of a leader, an acceptor,
        and a matchmaker. The first solid vertical black line shows the
        recovery of the leader, the second black line shows a reconfiguration
        away from the failed acceptor, and the third black line shows a
        reconfiguration away from the failed matchmaker.
      }
    }\figlabel{Chaos}
  \end{figure}
}{}

\begin{techreport}

  \textbf{Summary.}
  This experiment confirms that the extra latency of the Matchmaker phase during
  a leader change is negligible.
\end{techreport}

\subsection{Matchmaker Reconfiguration}\seclabel{EvalMatchmakerReconfiguration}
\textbf{Experiment Description.}
We deploy Matchmaker MultiPaxos as above. We again run three benchmarks with 1,
4, and 8 clients. Each benchmark runs for 40 seconds. During the first 10
seconds, there are no reconfigurations and no failures. Between 10 and 20
seconds, the leader reconfigures the set of matchmakers once every second.
Every reconfiguration randomly selects $2f+1$ matchmakers from a set of
$2\times(2f+1)$ matchmakers.
%
%
At 25 seconds, we fail a matchmaker. At 30 we perform a matchmaker
reconfiguration to replace the failed matchmaker. At 35 seconds, we reconfigure
the acceptors.
%

\begin{figure}[ht]
  \centering
  \includegraphics[width=\columnwidth]{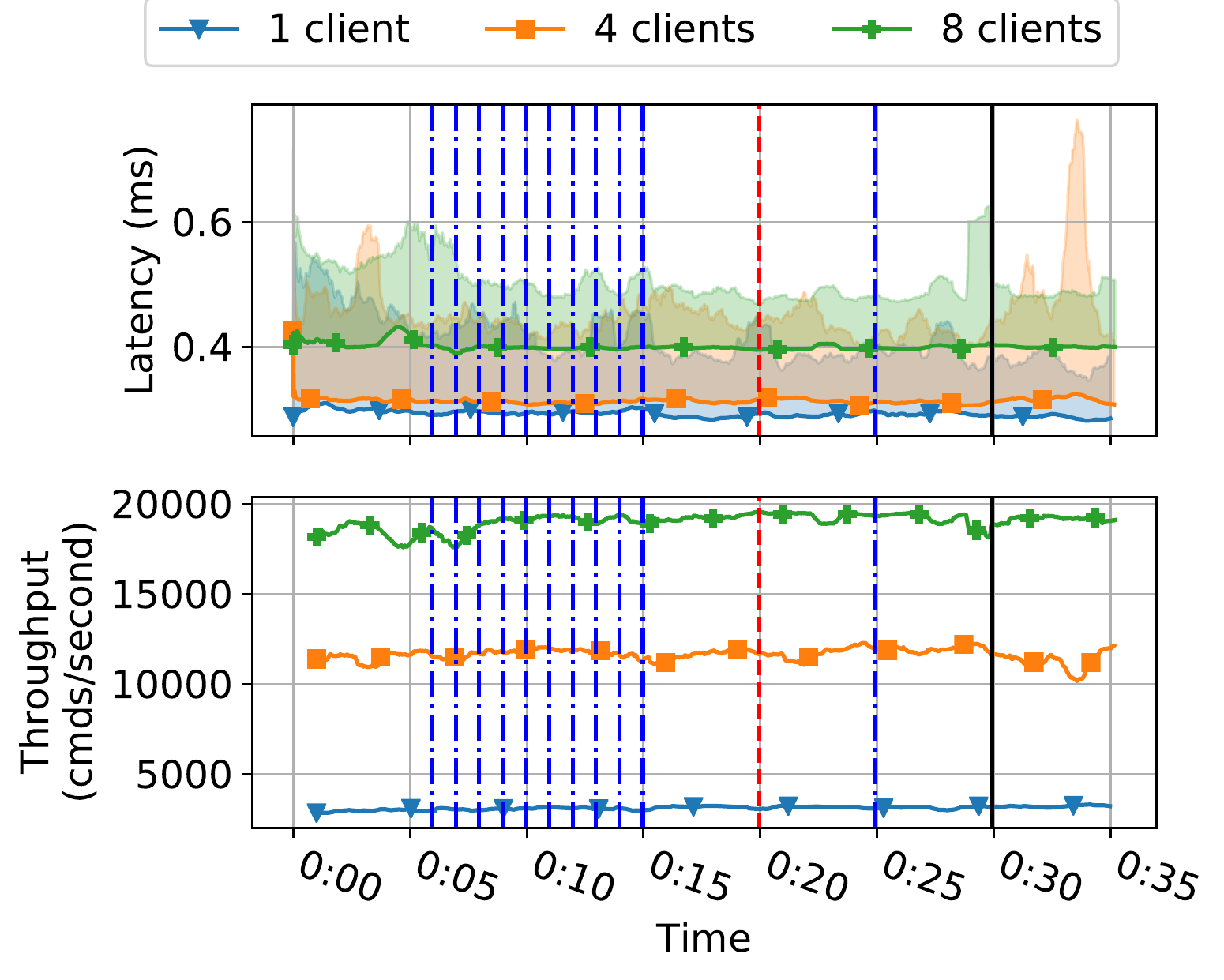}
  \caption{
    The latency and throughput of Matchmaker MultiPaxos ($f=1$). The
    dotted blue, dashed red, and vertical black lines show matchmaker
    reconfigurations, a matchmaker failure, and an acceptor reconfiguration
    respectively.
  }\figlabel{MatchmakerReconfigurationF1}
\end{figure}


\iftoggle{techreportenabled}{%
  {\begin{table}[ht]
  \centering
  \caption{%
   \figref{MatchmakerReconfigurationF1} median, interquartile range,
   and standard deviation of latency and throughput.
  }\tablabel{MatchmakerReconfigurationValues}
  Latency (ms)
  \begin{tabular}{l@{\hskip 4pt}c@{\hskip 4pt}cc@{\hskip 4pt}cc@{\hskip 4pt}c}
    \toprule
    & \multicolumn{2}{c}{1 Client}
    & \multicolumn{2}{c}{4 Clients}
    & \multicolumn{2}{c}{8 Clients} \\
    & 0s-10s & 10s-20s & 0s-10s & 10s-20s & 0s-10s & 10s-20s \\
    \cmidrule(lr){2-3}
    \cmidrule(lr){4-5}
    \cmidrule(lr){6-7}
    median & 0.297 & 0.292 & 0.314 & 0.313 & 0.404 & 0.398 \\
    IQR    & 0.032 & 0.024 & 0.031 & 0.030 & 0.035 & 0.028 \\
    stdev  & 0.077 & 0.061 & 0.093 & 0.098 & 0.383 & 0.067 \\
    \bottomrule
  \end{tabular}

  \vspace{12pt}
  Throughput (commands/second)
  \begin{tabular}{l@{\hskip 4pt}c@{\hskip 4pt}cc@{\hskip 4pt}cc@{\hskip 4pt}c}
    \toprule
    & \multicolumn{2}{c}{1 Client}
    & \multicolumn{2}{c}{4 Clients}
    & \multicolumn{2}{c}{8 Clients} \\
    & 0s-10s & 10s-20s & 0s-10s & 10s-20s & 0s-10s & 10s-20s \\
    \cmidrule(lr){2-3}
    \cmidrule(lr){4-5}
    \cmidrule(lr){6-7}
    median & 3019 & 3147 & 11631 & 11726 & 18569 & 19248 \\
    IQR    & 41   & 51   & 140   & 145   & 391   & 71 \\
    stdev  & 66   & 72   & 250   & 231   & 478   & 159 \\
    \bottomrule
  \end{tabular}
\end{table}
}
}{}

\textbf{Results.}
The latency and throughput of Matchmaker MultiPaxos are shown in
\figref{MatchmakerReconfigurationF1}. The latency and throughput of the
protocol remain steady through the first ten matchmaker reconfigurations,
through the matchmaker failure and recovery, and through the acceptor
reconfiguration. The medians, IQRs, and standard deviations of the latency and
throughput are very similar to the ones in
\tabref{LeaderReconfigurationValues}.

\begin{techreport}

  This is confirmed by the medians, IQRs, and standard deviations of the
  latency and throughput during the first 10 seconds and between 10 and 20
  seconds, which are shown in \tabref{MatchmakerReconfigurationValues}.
\end{techreport}

\begin{techreport}

  \textbf{Summary.}
  This benchmark confirms that matchmakers are off the critical path. The
  latency and throughput of Matchmaker MultiPaxos remains steady during a
  matchmaker reconfiguration and matchmaker failure. Moreover, a matchmaker
  reconfiguration does not affect the performance of subsequent acceptor
  reconfigurations.
\end{techreport}
}
{\section{Related Work}
\begin{techreport}

  \textbf{MultiPaxos' Horizontal Reconfiguration.}
  MultiPaxos' horizontal reconfiguration protocol~\cite{lamport1998part,
  lamport2010reconfiguring} uses the consensus it implements in order to reach
  consensus on a given configuration. This approach, also taken by systems such
  as Chubby~\cite{burrows2006chubby,chandra2007paxos}, does not require an
  extra reconfiguration algorithm or any additional nodes.
  Horizontal reconfiguration limits concurrency, as the proposer cannot have
  more than $\alpha$ outstanding operations. It can also be slow to
  reconfigure, as it needs to wait for $\alpha$ operations to be decided
  (though no-ops can help mitigate this). Unlike with Matchmaker Paxos,
  MultiPaxos cannot perform a reconfiguration if a proposer cannot contact a
  Phase 2 quorum of acceptors.
  Both MultiPaxos and Matchmaker MultiPaxos can perform a reconfiguration
  without disrupting the throughput or latency of the state machine.

  Matchmaker Paxos instead decouples reconfiguration from command processing.
  Matchmakers are rarely used and can be optimized for reliability, unlike the
  acceptors which are under high-load and thus optimized for performance. The
  approach of separating configuration management from the distributed system
  being configured is a common pattern and is also used by Chain
  Replication~\cite{van2004chain}.
\end{techreport}

\begin{techreport}

\textbf{SMART.}
  SMART~\cite{lorch2006smart} is a reconfiguration protocol that resolves many
  ambiguities in MultiPaxos' horizontal approach (e.g., when it is safe to
  retire old configurations). Like MultiPaxos' horizontal reconfiguration
  protocol, SMART can reconfigure a protocol with minimal performance
  degradation.

  SMART differs from Matchmaker Paxos is a number of ways. First, like
  MultiPaxos' horizontal reconfiguration protocol, SMART is fundamentally log
  based and is therefore incompatible with many sophisticated state machine
  replication protocols. Second, SMART assumes that acceptors and replicas are
  always co-located. This prevents us from reconfiguring the acceptors without
  reconfiguring the replicas. This is not ideal since we can reconfigure an
  acceptor without copying any state, but must transfer logs from an old
  replica to a new replica. SMART's garbage collection also has higher latency
  that Matchmaker Paxos' garbage collection. For Scenario 3, Matchmaker Paxos
  proposers wait until a prefix of the log is stored on $f+1$ replicas. SMART
  waits for the prefix of the log to be executed and snapshotted by $f+1$
  replicas.
\end{techreport}

\textbf{Raft.}
Raft~\cite{ongaro2014search} uses a reconfiguration protocol called joint
consensus. Like MultiPaxos' horizontal reconfiguration, joint consensus is
log-based and therefore incompatible with many existing replication protocols.

\begin{techreport}

  A simpler reconfiguration protocol for Raft was proposed
  in~\cite{ongaro2014consensus} but requires more rounds of communication.
\end{techreport}

\textbf{Viewstamped Replication (VR).}
VR~\cite{liskov2012viewstamped} uses a stop-the-world approach to
reconfiguration. During a reconfiguration, the entire protocol stops processing
commands. Thus, while the reconfiguration is quite simple, it is inefficient.
Stoppable Paxos~\cite{lamport2008stoppable} is similar to MultiPaxos'
horizontal reconfiguration, but also uses a stop-the-world approach.

\begin{techreport}

VR's stop-the-world approach to reconfiguration is also adopted by databases
built on VR, including TAPIR~\cite{zhang2018building} and
Meerkat~\cite{szekeres20meerkat}. We use a similar approach to reconfigure
matchmakers, but because matchmakers are off the critical path, the performance
overheads are invisible.
\end{techreport}

\textbf{Fast Paxos Coordinated Recovery.}
Fast Paxos has an optimization called coordinated recovery that is similar to
Phase 1 Bypassing. The main difference is that in coordinated recovery, a
leader uses Phase 2 information in round $i$ to skip Phase 1 in round $i+1$,
whereas with Phase 1 Bypassing, the leader instead uses Phase 1 information.

\begin{techreport}

  Note that coordinated recovery is not useful for Matchmaker MultiPaxos. It is
  subsumed by Phase 1 Bypassing. Coordinated recovery is only needed for Fast
  Paxos because the leader may not know which values were proposed in a round
  it owns. Phase 1 Bypassing cannot be applied to Fast Paxos for pretty much
  the same reason.
\end{techreport}

\begin{techreport}

  \textbf{DynaStore.}
  Vertical Paxos assumes its external master is implemented using state machine
  replication. MultiPaxos' horizontal reconfiguration also depends on
  consensus. Matchmaker Paxos does not require consensus to implement
  matchmakers, but we are not the first to notice this.
  DynaStore~\cite{aguilera2011dynamic} showed that reconfiguring atomic storage
  does not require consensus.
\end{techreport}

\textbf{ZooKeeper.}
ZooKeeper, a distributed coordinated service, which uses ZooKeeper Atomic
Broadcast~\cite{junqueira2011zab} is a protocol similar to MultiPaxos that can
also reconfigure quickly after leader failures.
}
{\section{Conclusion}
We presented Matchmaker Paxos and Matchmaker MultiPaxos to address the lack of
research on the increasingly important topic of reconfiguration. Our protocols
achieve a number of desirable properties, both theoretical and practical: they
can reconfigure without performance degradation, they provide insights into
existing protocols, and they generalize better than existing techniques.

}

\balance

\bibliographystyle{abbrv}
\bibliography{references}
\end{document}